\documentclass[aps,prc,nofootinbib,superscriptaddress,twocolumn,10pt,floatfix]{revtex4-2}

\usepackage{emnorm2}
\usepackage{writelabel}

\ifproofpre{}{\ifeof18
    \typeout{Version control information not updated!}
\else
    \immediate\write18{vc/vc}
\fi

\RequirePackage{xcolor}

\IfFileExists{vc/vc-info.tex}{\gdef\GITAbrHash{f751bf5}\gdef\VCRevision{\GITAbrHash}\gdef\VCDateText{July 22, 2025}\gdef\VCRevisionMod{\VCRevision}\gdef\VCRevisionMod{\VCRevision} }{
\gdef\GITAbrHash{unknown}\gdef\VCRevision{\GITAbrHash}\gdef\VCDateText{\today}\gdef\VCRevisionMod{\VCRevision}}
 }

\begin{document}

\title{Robust \textit{ab initio} predictions for dimensionless ratios of $E2$ and radius observables.
  II.~Estimation of $E2$ transition strengths by calibration to the charge radius}

\author{Mark A.~Caprio\,\orcidlink{0000-0001-5138-3740}}
\affiliation{Department of Physics and Astronomy, University of Notre Dame, Notre Dame, Indiana 46556-5670, USA}

\author{Patrick J.~Fasano\,\orcidlink{0000-0003-2457-4976}}
\altaffiliation[Present address: ]{NextSilicon Inc., Minneapolis, Minnesota 55402-1572, USA}
\affiliation{Department of Physics and Astronomy, University of Notre Dame, Notre Dame, Indiana 46556-5670, USA}
\affiliation{Physics Division, Argonne National Laboratory, Argonne, Illinois 60439-4801, USA}

\author{Pieter Maris\,\orcidlink{0000-0002-1351-7098}}
\affiliation{Department of Physics and Astronomy, Iowa State University, Ames, Iowa 50011-3160, USA}

\date{\ifproofpre{\today}{\VCDateText}}

\begin{abstract}
  Converged results for $E2$ observables are notoriously challenging to obtain in
\textit{ab initio} no-core configuration interaction (NCCI) approaches.  Matrix
elements of the $E2$ operator are sensitive to the large-distance tails of the
nuclear wave function, which converge slowly in an oscillator basis expansion.
Similar convergence challenges beset \textit{ab initio} prediction of the
nuclear charge radius.  However, we exploit systematic correlations between the
calculated $E2$ and radius observables to yield meaningful predictions for
relations among these observables.  In particular, we examine \textit{ab initio}
predictions for dimensionless ratios of the form $B(E2)/(e^2r^4)$, for nuclei
throughout the $p$ shell.  Meaningful predictions for $E2$ transition strengths
may then be made by calibrating to the ground-state charge radius, if
experimentally known.
 \end{abstract}

\ifproofpre{}{\preprint{Git hash: \VCRevisionMod}}

\maketitle

\writelabel{part1:sec:background}{II}
\writelabel{part1:sec:background:ratios}{II.1}
\writelabel{part1:sec:background:rc}{II.2}
\writelabel{part1:sec:background:deformation}{II.3}
\writelabel{part1:sec:results-moments:9be}{III}
\writelabel{part1:sec:results-moments:survey}{IV}
\writelabel{part1:sec:results-moments:deformation}{V}
\writelabel{part1:sec:app-geometric}{A}

\writelabel{part1:fig:q-norm-rp-scan-9be}{2}
\writelabel{part1:fig:q-norm-rp-scan-9be-convergence-diagnostics}{3}
\writelabel{part1:fig:q-norm-rp-ratio-teardrop-mirror}{4}
\writelabel{part1:fig:q-norm-rp-ratio-teardrop-neqz}{5}
\writelabel{part1:fig:q-norm-rp-ratio-teardrop-mirror-nrich-n}{6}
\writelabel{part1:fig:q-scan-6li}{7}
\writelabel{part1:fig:q-norm-rp-scan-9li}{8}
\writelabel{part1:fig:q-norm-rp-scan-8b-12b}{9}
\writelabel{part1:fig:beta-from-ratio-q-rsqr-scan-9be}{10}
\writelabel{part1:fig:beta-from-ratio-q-rsqr-teardrop-b}{11}

\writelabel{part1:eqn:ratio-rtp-q}{1}
\writelabel{part1:eqn:ratio-q-r}{2}
\writelabel{part1:eqn:ratio-rtp-r}{3}
\writelabel{part1:eqn:rc}{4}
\writelabel{part1:eqn:Q0-as-me}{5}
\writelabel{part1:eqn:rotational-e2-rme}{6}
\writelabel{part1:eqn:beta-as-norm}{7}
\writelabel{part1:eqn:Qmom-defn}{8}
\writelabel{part1:eqn:Q2mu-M0-beta-undifferentiated}{9}
\writelabel{part1:eqn:Q0-as-norm}{10}
\writelabel{part1:eqn:Q0-M0-beta-undifferentiated}{11}
\writelabel{part1:eqn:Q0-r-beta-undifferentiated}{12}
\writelabel{part1:eqn:Q-norm-r-beta-proton}{13}
\writelabel{part1:eqn:be2-norm-rp-beta-proton}{14}
\writelabel{part1:eqn:beta-be2-norm-r-proton-k0}{15}
\writelabel{part1:eqn:beta-be2-norm-r-proton-k0-traditional}{16}
\writelabel{part1:eqn:relative-difference}{17}
\writelabel{part1:eqn:difference-ratio}{18}
\writelabel{part1:eqn:difference-ratio-from-delta}{19}
\writelabel{part1:eqn:difference}{20}
\writelabel{part1:eqn:exp-convergence}{21}
\writelabel{part1:eqn:generalized-zeno-ratio}{A1}
\writelabel{part1:eqn:three-point-extrapolation}{A2}

\writelabel{part1:fn:q-be2-rme}{1}
\writelabel{part1:fn:e2-intrinsic}{2}
\writelabel{part1:fn:rot-reln}{3}

\section{Introduction}
\label{sec:intro}

Converged predictions for nuclear electric quadrupole ($E2$)
transition strengths are challenging to obtain in \textit{ab initio}
approaches~\cite{pervin2007:qmc-matrix-elements-a6-7,bogner2008:ncsm-converg-2N,maris2013:ncsm-pshell,carlson2015:qmc-nuclear,odell2016:ir-extrap-quadrupole,roth2023:em-properties-nuclei}.
Long-range observables, \textit{i.e.}, those which are sensitive to the
large-distance tails of the nuclear wave function, such as $E2$ matrix elements,
are slowly convergent in \textit{ab initio} no-core configuration interaction
(NCCI), or no-core shell-model (NCSM), calculations~\cite{barrett2013:ncsm}, as
these tails are described only with difficulty in an oscillator-basis expansion.

Nonetheless, one may exploit systematic correlations among calculated
observables to yield meaningful predictions for relations among these
observables, even where the observables individually are not adequately
converged (\textit{e.g.},
Refs.~\cite{calci2016:observable-correlations-chiral,sargsyan2022:8li-clustering-beta-recoil}).
Perhaps most naturally, the convergence patterns of different calculated $E2$
matrix elements may be strongly correlated with each other (\textit{e.g.},
Refs.~\cite{caprio2013:berotor,maris2015:berotor2,*maris2019:berotor2-ERRATUM,calci2016:observable-correlations-chiral,henderson2019:7be-coulex,caprio2021:emratio}).
In particular, the dimensionless ratio $B(E2)/(eQ)^2$, relating the $E2$
strength to the quadrupole moment, is in many cases found to be robustly
convergent~\cite{caprio2013:berotor,maris2015:berotor2,*maris2019:berotor2-ERRATUM,maris2013:ncci-chiral-ccp12,calci2016:observable-correlations-chiral,dalessio2020:12c-escatt-be2,caprio2022:8li-trans,li2024:quadrupole-10be-c}.
A meaningful prediction of the $E2$ strength may then be obtained by calibration
to a measured ground-state quadrupole moment~\cite{caprio2022:8li-trans}.

\begin{figure}[b]
\begin{center}
\includegraphics[width=\ifproofpre{1}{1}\hsize]{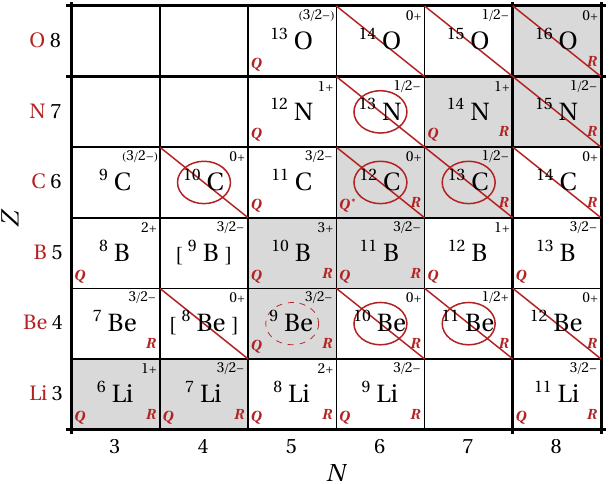}
\end{center}
\caption{Overview of particle-bound nuclides in the $p$ shell, highlighting
  nuclides for which $E2$ strengths are considered, in relation to the radius,
  in this work: the nuclide $\isotope[9]{Be}$ (dashed circle), for which the ground-state angular momentum \textit{does}
  also support a quadrupole moment, and the remaining nuclides (solid circle), for which it does
  \textit{not}.  Nuclides with measured ground-state quadrupole
  moments~\cite{stone2016:e2-moments} and charge
  radii~\cite{angeli2013:charge-radii,npa2017:012} are indicated by the letter
  ``$Q$'' or ``$R$'', respectively, while a measured excited-state quadrupole
  moment~\cite{stone2016:e2-moments} is indicated by ``$Q^*$''.  Brackets
  indicate a particle-unbound but narrow ($\lesssim 1\,\keV$) ground-state
  resonance, shading indicates beta-stable nuclides, and the experimental
  ground-state angular momentum and parity are
  given~\cite{npa2002:005-007,npa2004:008-010,npa2012:011,npa2017:012,npa1991:013-015}.
  Nuclei for which the ground-state angular momentum does not support a
  quadrupole moment ($J\leq1/2$) are crossed out with a diagonal line.
}
\label{fig:nuclear-chart-be2}
\end{figure}

However, correlations among the $E2$ matrix elements are not the only ones which
might be exploited.  There are cases in which the ground-state
quadrupole moment is not suitable for use in calibration, either since it is
unknown, or since it is subject to large calculational uncertainties and/or
sensitivity to the internucleon interaction (as for the suppressed quadrupole
moment in
$\isotope[6]{Li}$~\cite{cockrell2012:li-ncfc,calci2016:observable-correlations-chiral}),
or, more simply, since only states with angular momentum $J\geq1$ admit a
nonvanishing quadrupole moment.  Thus, notably, calibration to the ground-state
quadrupole moment is not possible for even-even nuclei, nor for odd-mass nuclei
with $J=1/2$ ground states.  (On a nuclear chart of the $p$ shell, in
Fig.~\ref{fig:nuclear-chart-be2}, such nuclides are crossed out with a diagonal
line.)
  
The electric monopole ($E0$) and $E2$ operators have a similar sensitivity
($\propto r^2$) to the large-distance behavior of the nucleons in the nuclear
many-body wave function.  Therefore, we might hope to find $E2$ observables also
to be correlated with the $E0$ moment or, equivalently, root mean square
(r.m.s.) radius.  If the dimensionless ratio $B(E2)/(e^2r^4)$, now involving the
ground-state radius, is robustly convergent, then a meaningful prediction of the
$E2$ strength may again be obtained, now by calibration to a measured ground
state (charge) radius.

Notionally, at least, these different dimensionless ratios, $B(E2)/(eQ)^2$
and $B(E2)/(e^2r^4)$, are sensitive to different aspects of nuclear
structure.  In simple model limits~\cite{casten2000:ns}, the ratio $B(E2)/(eQ)^2$
is uniquely determined by symmetry considerations.  \textit{E.g.}, taking the
axially symmetric rotor as a principal example, $B(E2)/(eQ)^2$ follows from the
rotational relations among electromagnetic matrix
elements~\cite{rowe2010:collective-motion}, independent of the magnitude of the
nuclear quadrupole deformation.  The ratio $B(E2)/(e^2r^4)$, instead, provides a measure of
this deformation.  In this same rotational picture, it is proportional to the
square of the Bohr deformation variable $\beta$~\cite{bohr1952:vibcoupling}.

In a previous article (Part~I)~\cite{emnorm2-part1}, we considered the relation
between calculated $E2$ and $E0$ moments in the ground state, namely, through
the dimensionless ratio $Q/r^2$.  We did so for nuclei across the $p$ shell.
(If $Q/r^2$ is robustly convergent, then a meaningful prediction of the
quadrupole moment may be obtained, by calibration to a measured ground state
radius, or \textit{vice versa}.)  The reader is also referred to Part~I for
background on the relation of such ratios to the quadrupole deformation, as well
as the distinction between the point-proton radius $r_p$ considered in nuclear
structure calculations and the measurable charge radius $r_c$.

In the present article (Part~II), we address $E2$ transition strengths, through
the dimensionless ratio $B(E2)/(e^2r^4)$, for low-lying $E2$ transitions in
selected $p$-shell nuclei.  We focus primarily on nuclei for which the ground
state does not support a quadrupole moment (that is, nuclei for which
calibration to the ground-state radius is, at least in principle, possible,
while calibration to the quadrupole moment is not).

First, we compare the convergence properties of the dimensionless ratios
$B(E2)/(eQ)^2$ and $B(E2)/(e^2r^4)$, taking for illustration a nuclide for which
the ground-state angular momentum \textit{does} support a quadrupole moment,
namely, $\isotope[9]{Be}$ [Fig.~\ref{fig:nuclear-chart-be2} (dashed circle)]
(Sec.~\ref{sec:results-trans-9be}).  We then relate the $E2$ strength to the
radius, via $B(E2)/(e^2r^4)$, for transitions in $p$-shell nuclei for which the
ground-state angular momentum does \textit{not} support a quadrupole moment
(Sec.~\ref{sec:results-trans-survey}): namely, the even-even ($J=0$) nuclides
$\isotope[10]{Be}$, $\isotope[10]{C}$, and $\isotope[12]{C}$, and the odd-mass
($J=1/2$) nuclides $\isotope[11]{Be}$, $\isotope[13]{C}$, and $\isotope[13]{N}$
[Fig.~\ref{fig:nuclear-chart-be2} (solid circles)].  We also consider
dimensionless ratios $B(E2)/(eQ)^2$, involving the quadrupole moment of the
excited initial state for the transition, as a diagnostic for axially symmetric
rotational structure, as well as $Q/r^2$ for the experimental measured $2^+$
quadrupole moment of
$\isotope[12]{C}$~\cite{stone2016:e2-moments,vermeer1983:12c-coulex,saizlomas2023:12c-coulex}.
Finally, we use these results to extract the deformations of the proton and
neutron distributions of the nuclear intrinsic state, under the assumption of
axial symmetry (Sec.~\ref{sec:results-trans-deformation}).  Preliminary
results were reported in Ref.~\cite{caprio2022:emnorm}.

 \section{Convergence illustration: Calibration to the quadrupole moment and radius in $\isotope[9]{Be}$}
\label{sec:results-trans-9be}

\begin{figure*}
\centering
\includegraphics[width=\ifproofpre{1.0}{1.0}\hsize]{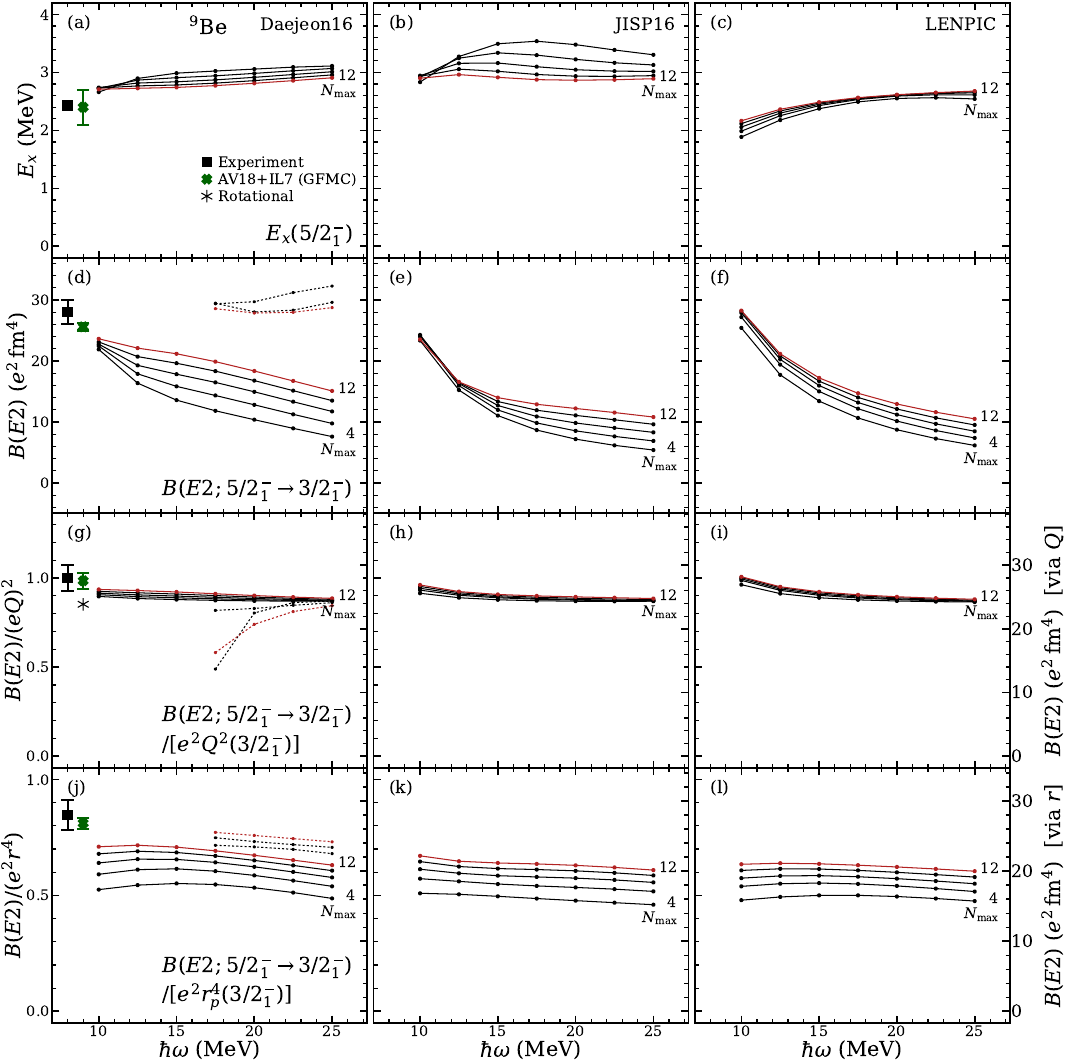}
\caption{Calculated transition observables for $\isotope[9]{Be}$:
$E_x(5/2^-)$,
$B(E2;5/2^-_1\rightarrow 3/2^-_1)$,
the dimensionless ratio $B(E2)/(eQ)^2$,
and the dimensionless ratio $B(E2)/(e^2r_p^4)$~(from top to bottom).
Results are shown for the Daejeon16 (left),
  JISP16 (center), and LENPIC (right) interactions.  Calculated values are shown
  as functions of the basis parameter $\hw$, for successive even values of
  $\Nmax$, from $\Nmax=4$ to $12$ (as labeled).  When calibrated to the
  experimentally deduced value for $Q$ or $r_p$, the ratio provides a prediction
  for the absolute $B(E2)$ (scale at right).  Exponential extrapolations (small
  circles, dotted lines) are provided, for selected observables, for the
  Daejeon16 results only ($\hw\geq17.5\,\MeV$).  For comparison, experimental
  values~\cite{stone2016:e2-moments,angeli2013:charge-radii,npa2004:008-010}
  (squares), GFMC AV18+IL7 predictions~\cite{pastore2013:qmc-em-alt9} (see also
  Table~III of Ref.~\cite{carlson2015:qmc-nuclear} for $E_x$) (crosses), and the
  rotational $E2$ ratio (asterisk) are also shown.
  Includes results for $B(E2)/(eQ)^2$ previously shown (for $\Nmax\leq10$) in Ref.~\cite{caprio2022:8li-trans}.
}
\label{fig:be2-norm-comparison-scan-9be}
\end{figure*}

As an illustrative example, let us take the lowest lying $E2$ transition in
$\isotope[9]{Be}$, that is, the $5/2^-\rightarrow3/2^-$ transition from the
first negative parity excited state to the ground state.  We have already
examined the convergence of the ground state quadrupole moment and radius, and
the dimensionless ratio $Q/r^2$ involving these quantities, in
Sec.~\ref{part1:sec:results-moments:9be} of Part~I.  Although the $5/2^-$ state,
at $2.42\,\MeV$, lies above the neutron threshold, it is a narrow resonance,
with a width of $\approx0.8\,\keV$~\cite{npa2004:008-010}.  The $3/2^-$ ground
state may be interpreted as the band head of a $K=3/2$ rotational
band~\cite{millener2001:light-nuclei,millener2007:p-shell-hypernuclei,caprio2020:bebands},
and the $5/2^-$ state as a band member, so that the $5/2^-\rightarrow 3/2^-$
transition is a rotational in-band transition (see Fig.~1 of
Ref.~\cite{caprio2020:bebands}).  The $E2$ strength for this transition is
measured experimentally as $B(E2;5/2^-\rightarrow
3/2^-)=28(2)\,e^2\fm^4$~\cite{npa2004:008-010}.

Calculated observables pertaining to this $5/2^-\rightarrow 3/2^-$ transition in
$\isotope[9]{Be}$ are shown in Fig.~\ref{fig:be2-norm-comparison-scan-9be}.
Each curve in Fig.~\ref{fig:be2-norm-comparison-scan-9be} represents the results
of calculations sharing the same truncation $\Nmax$, for the many-body harmonic
oscillator basis, but with varying choices of the oscillator scale parameter
$\hw$~\cite{barrett2013:ncsm}, as described in Part~I.  Calculations are carried
out using the NCCI code
MFDn~\cite{maris2010:ncsm-mfdn-iccs10,*shao2018:ncci-preconditioned} and the
associated postprocessor code for transitions~\cite{fasano2023:diss}.

In Fig.~\ref{fig:be2-norm-comparison-scan-9be}, we take NCCI calculations based
on three different internucleon interactions (left to right), as described in Part~I:
Daejeon16~\cite{shirokov2016:nn-daejeon16},
JISP16~\cite{shirokov2007:nn-jisp16}, and a chiral effective field theory
($\chi$EFT) interaction, namely, the two-body part of the \ntwolo{} LENPIC
interaction (with semi-local coordinate-space regulator parameter
$R=1\,\fm$)~\cite{epelbaum2015:lenpic-n4lo-scs,epelbaum2015:lenpic-n3lo-scs}.
Of these, the Daejeon16 interaction is the ``softest'', providing the most
favorable convergence properties, while the LENPIC interaction is the
``hardest'', since here it is taken with no similarity renormalization group
(SRG) softening.  Also shown in Fig.~\ref{fig:be2-norm-comparison-scan-9be} are
experimental values (squares) and the predictions of Green's function Monte Carlo
(GFMC) calculations~\cite{pastore2013:qmc-em-alt9} with the Argonne $v_{18}$
(AV18) two-nucleon~\cite{wiringa1995:nn-av18} and Illinois-7 (IL7)
three-nucleon~\cite{pieper2008:3n-il7-fm50} potentials (crosses).

The $5/2^-$ excitation energy (\textit{i.e.}, the energy difference for the
$5/2^-\rightarrow 3/2^-$ transition) is shown in
Fig.~\ref{fig:be2-norm-comparison-scan-9be} (first row).  We note rough
consistency in the scale of the excitation energy, both across the different
calculations, and with experiment.

\begin{figure}
\centering
\includegraphics[width=\ifproofpre{0.9}{0.45}\hsize]{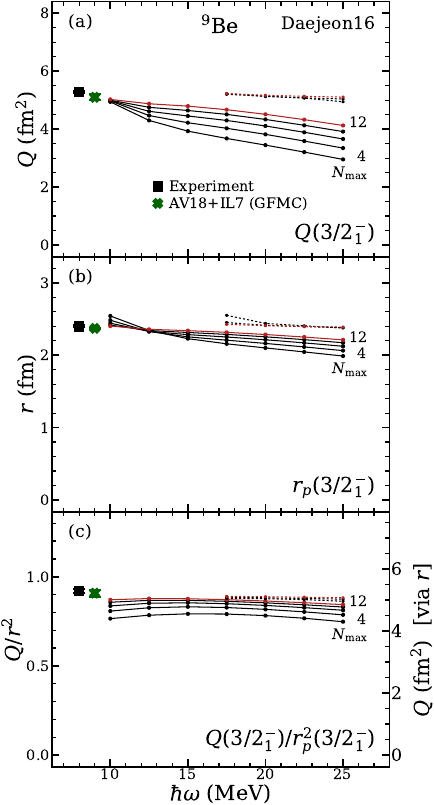}
\caption{Calculated ground state observables for $\isotope[9]{Be}$:
  (a)~$Q(3/2^-_1)$,
  (b)~$r_p(3/2^-_1)$,
  and
  (c)~the dimensionless ratio $Q/r_p^2$.
Results are shown for the Daejeon16 interaction.  Calculated values are shown
  as functions of the basis parameter $\hw$, for successive even values of
  $\Nmax$, from $\Nmax=4$ to $12$ (as labeled).  When calibrated to the
  experimentally deduced value for $r_p$, the ratio provides a prediction
  for the absolute $Q$ (scale at right).  Exponential extrapolations (small
  circles, dotted lines) are provided ($\hw\geq17.5\,\MeV$).  For comparison, experimental
  values~\cite{stone2016:e2-moments,angeli2013:charge-radii}
  (squares) and GFMC AV18+IL7 predictions~\cite{pastore2013:qmc-em-alt9} (crosses) are also shown.
  Results reproduced from Fig.~\ref{part1:fig:q-norm-rp-scan-9be} of Part~I.
}
\label{fig:q-norm-rp-scan-9be}
\end{figure}
\begin{figure*}
\centering
\includegraphics[width=\ifproofpre{0.9}{0.9}\hsize]{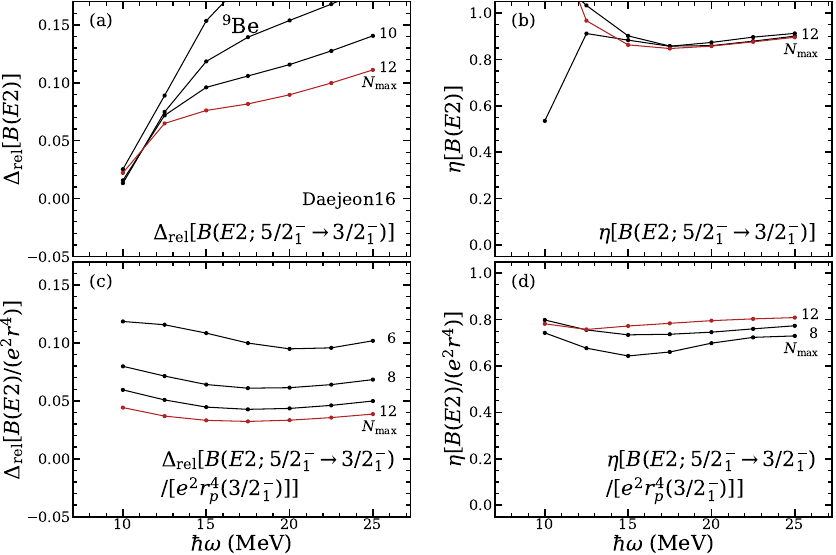}
\caption{Diagnostics of convergence for transition observables for
  $\isotope[9]{Be}$: the relative difference $\Deltarel$~(left) and ratio of
  successive differences $\eta$~(right), for $B(E2;5/2^-_1\rightarrow
  3/2^-_1)$~(top) and the dimensionless ratio $B(E2)/(e^2r_p^4)$~(bottom).
  Calculated values, for the Daejeon16 interaction, are shown as functions of
  the basis parameter $\hw$, for successive even values of $\Nmax$, from from
  $\Nmax=6$ or $8$ (as appropriate, given observables calculated starting with
  $\Nmax=4$) to $12$ (as labeled).  }
\label{fig:be2-norm-rp-scan-9be-convergence-diagnostics}
\end{figure*}

Before proceeding to the $E2$ transition strength, it is worth recalling from
Part~I the convergence behaviors of the calculated ground state quadrupole
moment $Q$ and proton radius $r_p$.  For the reader's convenience, we replicate
in Fig.~\ref{fig:q-norm-rp-scan-9be} illustrative results for $Q$, $r_p$, and
$Q/r_p^2$, for the Daejeon16 interaction [from
  Fig.~\ref{part1:fig:q-norm-rp-scan-9be}~(d,g,j) of Part~I].  For both $Q$
[Fig.~\ref{fig:q-norm-rp-scan-9be}(a)] and $r_p$
[Fig.~\ref{fig:q-norm-rp-scan-9be}(b)], we see a flattening (shouldering) of the
curves, in the lower portion of the $\hw$ range shown, and the spacing between
curves for successive $\Nmax$ decreases.  While this behavior suggests an
approach to convergence, the level of convergence is not sufficient for us to
read off a concrete estimate of the true result for the observable in the full,
untruncated space.  (For the harder JISP16 and LENPIC interactions, any such
hints of convergence, especially shouldering, are less obvious.)  For $Q$, the
curves appear to approximately cross at a single point at or near the low end of
the $\hw$ range shown, or, for $r_p$, at a point somewhat higher in $\hw$ (by
$\approx2.5\,\MeV$).  Then, taking the ratio $Q/r_p^2$
[Fig.~\ref{fig:q-norm-rp-scan-9be}(c)] removes much of the $\Nmax$ and $\hw$
dependence observed for $Q$, although some $\hw$ dependence remains
(specifically, for the Daejeon16 results, taking the form of a gentle fall-off
on either side of $\hw\approx15\,\MeV$).

Turning now to the present transition of interest, the calculated $B(E2;
5/2^-\rightarrow 3/2^-)$ is shown in
Fig.~\ref{fig:be2-norm-comparison-scan-9be}~(second row).  For the results
obtained with the Daejeon16 interaction
[Fig.~\ref{fig:be2-norm-comparison-scan-9be}(d)], this quantity likewise shows
some indication of shouldering in the lower portion of the $\hw$ range shown, as
well as some slight indication of compression with increasing $\Nmax$ in the
shouldering region.  Again, there is no way to read off the true, converged
value directly from these results with any confidence.  Regarding the other, harder
interactions, there is little sign of convergence (either compression of the
curves or shouldering) in the calculated $B(E2)$ for JISP16
[Fig.~\ref{fig:be2-norm-comparison-scan-9be}(e)] and essentially none for the
bare LENPIC interaction [Fig.~\ref{fig:be2-norm-comparison-scan-9be}(f)].

We begin our consideration of dimensionless ratios with $B(E2)/(eQ)^2$, a ratio
involving only $E2$ matrix elements, as proposed in
Ref.~\cite{caprio2022:8li-trans}.  The calculated results for this ratio are
shown for $\isotope[9]{Be}$ in Fig.~\ref{fig:be2-norm-comparison-scan-9be}
(third row).  (Calibrating to the experimental ground state quadrupole
moment~\cite{stone2016:e2-moments} gives the scale shown at right.)  The $\hw$
dependence seen in the $B(E2)$ [Fig.~\ref{fig:be2-norm-comparison-scan-9be}
  (second row)] is essentially eliminated in the ratio $B(E2)/(eQ)^2$, and
convergence with $\Nmax$ is remarkably rapid.  For the Daejeon16 results
[Fig.~\ref{fig:be2-norm-comparison-scan-9be}(g)], the values obtained for
successive $\Nmax$ differ by $\lesssim1.5\%$ even for the lowest $\Nmax$ shown.
These Daejeon16 results lie just below the experimental ratio (square) and GFMC
AV18+IL7 predictions~\cite{pastore2013:qmc-em-alt9} (cross) for the ratio, and
just above (by $\approx5\%$) the $K=3/2$ rotational value
[$B(E2)/(eQ)^2=75/(28\pi)\approx0.8526$] (asterisk).\footnote{All necessary
relations~\cite{rowe2010:collective-motion} for calculating the rotational ratio
may be found in footnotes~\ref{part1:fn:q-be2-rme} and~\ref{part1:fn:rot-reln}
of Part~I.}

However, our principal interest here is in the dimensionless ratio
$B(E2)/(e^2r_p^4)$, shown for $\isotope[9]{Be}$ in
Fig.~\ref{fig:be2-norm-comparison-scan-9be} (fourth row).  (Calibrating to the
experimental ground state point-proton radius~\cite{angeli2013:charge-radii}
gives the scale shown at right.)  Taking this ratio again serves to eliminate
much of the $\hw$ dependence found in the $B(E2)$ itself
[Fig.~\ref{fig:be2-norm-comparison-scan-9be}~(second row)].  For the Daejeon16
results [Fig.~\ref{fig:be2-norm-comparison-scan-9be}(j)], there is still a
modest $\hw$ dependence, similar to that seen above for $Q/r_p^2$
[Fig.~\ref{fig:q-norm-rp-scan-9be}(c)], though this is not seen in the results
for the other interactions [Fig.~\ref{fig:be2-norm-comparison-scan-9be}(k,l)].
Taking the ratio reduces, but does not eliminate, the $\Nmax$ dependence found
for the $B(E2)$ itself.  We may note that there is no apparent crossing point
for the $B(E2)$, with respect to $\Nmax$, in the $\hw$ range shown
[Fig.~\ref{fig:be2-norm-comparison-scan-9be}~(d)].  Therefore, it is already
clear that the $B(E2)$ cannot be \textit{strictly} correlated, in a one-to-one
relation, to either $Q$ [Fig.~\ref{fig:q-norm-rp-scan-9be}(a)] or $r_p$
[Fig.~\ref{fig:q-norm-rp-scan-9be}(b)].  What is most striking is the
improvement obtained, by taking the ratio, for the harder interactions
[Fig.~\ref{fig:be2-norm-comparison-scan-9be}(h,i)], where the underlying $B(E2)$
[Fig.~\ref{fig:be2-norm-comparison-scan-9be}(e,f)] is itself so poorly
converged.\footnote{Given the close proportionality between $B(E2)$ and $Q^2$
established above, note that the convergence properties of the ratio
$B(E2)/(e^2r_p^4)$ found here [Fig.~\ref{fig:be2-norm-comparison-scan-9be}
  (fourth row)] are largely foretold by those previously found (in Part~I) for
the ratio $Q/r_p^2$ [Fig.~\ref{part1:fig:q-norm-rp-scan-9be} (bottom row) of
  Part~I]. Since the numerator and denominator are now quadratic, rather than
linear, in the underlying matrix elements (of the monopole and quadrupole
operators), relative changes in $B(E2)/(e^2r_p^4)$ may be expected to be
amplified by a factor of $2$ compared to those in $Q/r_p^2$.}

At least in this example, the ratio $B(E2)/(e^2r_p^4)$ is not as robustly
convergent as $B(E2)/(eQ)^2$.  Nonetheless, it would still appear feasible to
obtain a rough estimate (or at least a lower bound) for the $B(E2)$ by
calibration to the ground-state radius, using the \textit{ab initio}
calculations for this ratio in Fig.~\ref{fig:be2-norm-comparison-scan-9be}
(fourth row).

As in Part~I, we may also consider quantitative measures of convergence, shown
in Fig.~\ref{fig:be2-norm-rp-scan-9be-convergence-diagnostics} for both the
$B(E2)$ itself [Fig.~\ref{fig:be2-norm-rp-scan-9be-convergence-diagnostics}
  (top)] and the dimensionless ratio $B(E2)/(e^2r_p^4)$
[Fig.~\ref{fig:be2-norm-rp-scan-9be-convergence-diagnostics} (bottom)].
Namely, we show the relative differences $\Deltarel$
[Fig.~\ref{fig:be2-norm-rp-scan-9be-convergence-diagnostics} (left)], defined
in~(\ref{part1:eqn:relative-difference}) of Part~I, and the ratio $\eta$ of of
successive differences
[Fig.~\ref{fig:be2-norm-rp-scan-9be-convergence-diagnostics} (right)], defined
in~(\ref{part1:eqn:difference-ratio-from-delta}) of Part~I.  For exponential
convergence~\cite{forssen2008:ncsm-sequences,bogner2008:ncsm-converg-2N}, that
is, $\Nmax$ dependence of the form [see (\ref{part1:eqn:exp-convergence}) of
  Part~I]
\begin{displaymath}
X(\Nmax)=X_{\infty}+a\exp(-c\Nmax),
\end{displaymath}
an observable $X$ approaches its limiting value in steps the sizes of which form
a geometric progression.  In this case, $\eta$ should be constant with respect
to $\Nmax$, at fixed $\hw$.  In Part~I, it was found (see
Fig.~\ref{part1:fig:q-norm-rp-scan-9be-convergence-diagnostics} of Part~I) that
the ground state observables $Q$ and $r_p$ exhibit roughly exponential convergence with respect to
$\Nmax$, in the $\hw$ regions safely above any crossing point, but this
convergence is slow, with $\eta\approx0.8$.

For the $B(E2)$, the convergence is again seen to be roughly exponential, for
the upper part of the $\hw$ range shown ($\hw\gtrsim15\,\MeV$), in that $\eta$
[Fig.~\ref{fig:be2-norm-rp-scan-9be-convergence-diagnostics}~(b)] is
approximately independent of $\Nmax$, at each $\hw$.  The convergence rate, as
measured by $\eta$, is weakly dependent on $\hw$.  However, it is generally even
slower ($\eta\approx0.9$) than for $Q$ and $r_p$.  For $B(E2)/(e^2r_p^4)$, the
relative differences
[Fig.~\ref{fig:be2-norm-rp-scan-9be-convergence-diagnostics}(c)] are no longer
highly $\hw$ dependent, as they are for the $B(E2)$ itself
[Fig.~\ref{fig:be2-norm-rp-scan-9be-convergence-diagnostics}(a)].  The ratio of
successive differences for $B(E2)/(e^2r_p^4)$
[Fig.~\ref{fig:be2-norm-rp-scan-9be-convergence-diagnostics}(d)] is somewhat
reduced ($0.6\lesssim\eta\lesssim0.8$) relative to those for the $B(E2)$ itself
[Fig.~\ref{fig:be2-norm-rp-scan-9be-convergence-diagnostics}~(b)], but this
ratio is now more dependent upon $\Nmax$, signaling a convergence pattern which
is less exponential.  As noted in Part~I, the ratio of two exponentially
converging quantities need not itself be exponentially converging (although it
may be approximately so in limiting cases).

If we assume exponential convergence, on a purely empirical basis, we may
extrapolate from values calculated at finite $\Nmax$ to the limiting value
$X_{\infty}$.  A simple three-point exponential extrapolation (see
Appendix~\ref{part1:sec:app-geometric} of Part~I) is given by the small circles
and dotted curves in Figs.~\ref{fig:be2-norm-comparison-scan-9be}
and~\ref{fig:q-norm-rp-scan-9be}, for the Daejeon16 results.  (From the
calculations considered here, with $\Nmax\geq4$, a three-point extrapolation
becomes possible for $\Nmax\geq8$.  Extrapolations are shown only in the $\hw$
region safely above any crossing point for the ground state $Q$ and $r_p$.)  The
extrapolations for $Q$ [Fig.~\ref{fig:q-norm-rp-scan-9be}~(a)] and $r_p$
[Fig.~\ref{fig:q-norm-rp-scan-9be}~(b)], for the Daejeon16 results, were seen in
Part~I to be remarkably robust, \textit{i.e.}, independent of $\Nmax$ and $\hw$.
The extrapolated values are also consistent with both experiment (squares) and
the GFMC AV18+IL7 results~\cite{pastore2013:qmc-em-alt9} (crosses).  However,
for the harder JISP16 and LENPIC interactions, three-point exponential
extrapolation yields results which are significantly less robust, and thus
extrapolations for these interactions are not shown in
Figs.~\ref{fig:be2-norm-comparison-scan-9be} or~\ref{fig:q-norm-rp-scan-9be}
(but may be found in the Supplemental Material~\cite{supplemental-material}).

We can again consider the results of a simple three-point exponential
extrapolation for the $B(E2)$ [Fig.~\ref{fig:be2-norm-comparison-scan-9be}~(d)].
For the two highest $\Nmax$ values, the extrapolations become largely consistent
and $\hw$ independent (over the $\hw$ range shown), suggesting that
extrapolation might be feasible.  These extrapolations are also roughly
consistent with experiment (square) and the GFMC AV18+IL7
result~\cite{pastore2013:qmc-em-alt9} (cross).  Note that exponential
extrapolation does not appear to be helpful for the ratio $B(E2)/(eQ)^2$
[Fig.~\ref{fig:be2-norm-comparison-scan-9be}~(g)], not unexpectedly, since the
small changes with $\Nmax$ would not appear, even by eye, to follow a simple
exponential convergence pattern.  Then, as noted above, the convergence of
$B(E2)/(e^2r_p^4)$ is less exponential, as indicated by $\eta$
[Fig.~\ref{fig:be2-norm-rp-scan-9be-convergence-diagnostics} (right)], than for
the $B(E2)$ itself.  Indeed, simple three-point exponential extrapolation
[Fig.~\ref{fig:be2-norm-comparison-scan-9be}(j)] does not yield a consistent
value.

\section{$E2$ transition strengths by calibration to the ground state radius in $p$-shell nuclei}
\label{sec:results-trans-survey}

\begin{figure*}
\centering
\includegraphics[width=\ifproofpre{0.6}{0.75}\hsize]{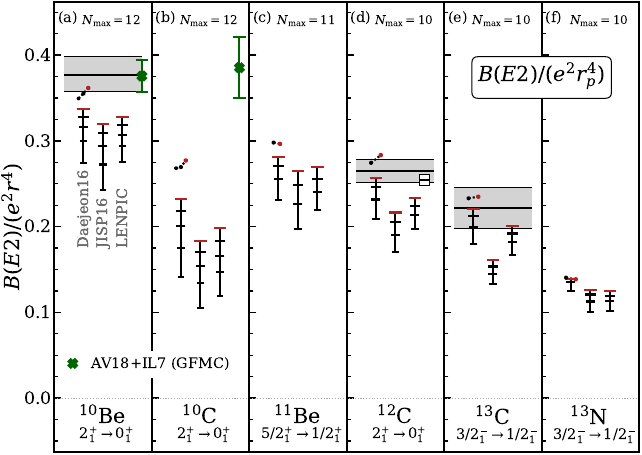}
\caption{Calculated ratios $B(E2)/(e^2r_p^4)$, for transitions involving the
  ground states of nuclides in the $p$ shell, considering cases in which the
  ground-state angular momentum does \textit{not} support a quadrupole moment,
  namely, $\isotope[10]{Be}$, $\isotope[10]{C}$, $\isotope[11]{Be}$,
  $\isotope[12]{C}$, $\isotope[13]{C}$, and $\isotope[13]{N}$ (left to right).
  Results are obtained with the Daejeon16, JISP16, and LENPIC interactions (from
  left to right, within each panel).  Exponential extrapolations for
  $B(E2)/(e^2r_p^4)$ (small circles, dotted lines) are provided, for the
  Daejeon16 results only (plotted with $\Nmax$ increasing from left to right).
  For comparison, the experimental
  ratios~\cite{angeli2013:charge-radii,pritychenko2016:e2-systematics,npa1991:013-015}
  are shown (horizontal line and error band), as are the GFMC AV18+IL7
  predictions~\cite{mccutchan2012:10c-dsam,carlson2015:qmc-nuclear} for $A=10$
  (crosses).  For $\isotope[12]{C}$, the ratio deduced from the more recent
  $B(E2)$ measurement of D'Alessio \textit{et
    al.}~\cite{dalessio2020:12c-escatt-be2} is also included (horizontal line
  and error band, shown as narrow open box).  }
\label{fig:be2-norm-rp-ratio-teardrop-forbidden-q}
\end{figure*}

\begin{figure*}
\centering
\includegraphics[width=\ifproofpre{0.9}{1.0}\hsize]{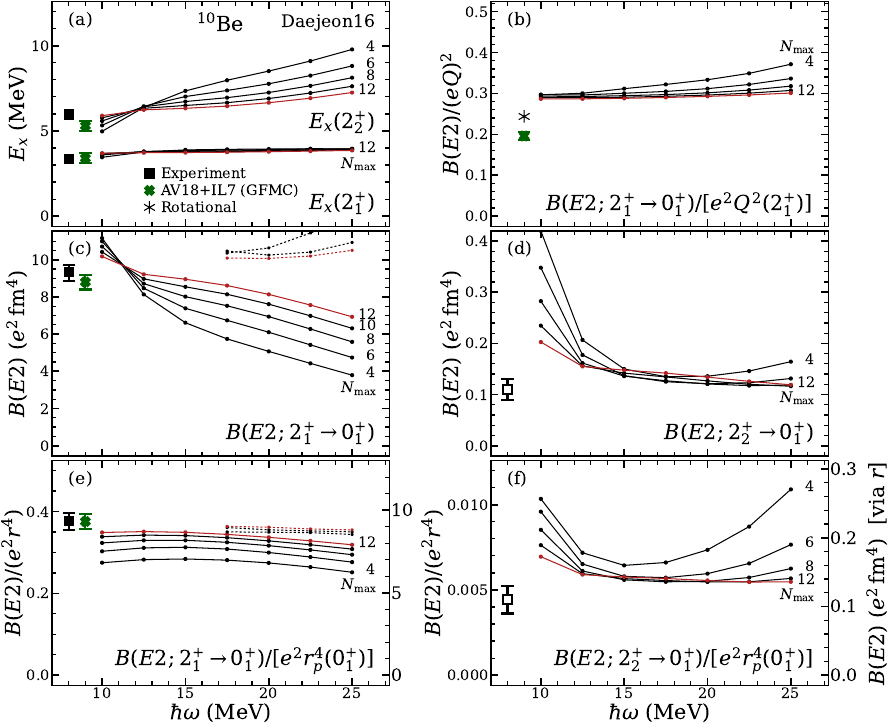}
\caption{Calculated observables for $\isotope[10]{Be}$: (a)~the $2^+_1$ and
  $2^+_2$ excitation energies, (b)~the dimensionless ratio $B(E2)/(eQ)^2$, for
  the transition from the $2^+_1$ state and taken relative to the quadrupole
  moment of that state, (c,d) the $E2$ strengths $B(E2; 2^+_1\rightarrow 0^+_1)$
  and $B(E2; 2^+_2\rightarrow 0^+_1)$, and (e,f)~the corresponding dimensionless
  ratios $B(E2)/(e^2r_p^4)$, taken relative to the ground state
  radius. Calculated values, for the Daejeon16 interaction, are shown as
  functions of the basis parameter $\hw$, for successive even values of $\Nmax$,
  from $\Nmax=4$ to $12$ (as labeled).  When calibrated to the experimentally
  deduced value for $r_p$, the ratio provides a prediction for the absolute
  $B(E2)$ (scale at right).  Exponential extrapolations (small circles, dotted
  lines) are provided, for selected observables.  For comparison, the
  experimental
  values~\cite{npa2004:008-010,angeli2013:charge-radii,pritychenko2016:e2-systematics},
  including evaluated results (solid squares) and the recent first measurement
  of McCutchan~\textit{et al.}~\cite{mccutchan2009:10be-dsam-gfmc} for the
  $2^+_2\rightarrow 0^+_1$ strength (open squares), GFMC AV18+IL7
  predictions~\cite{mccutchan2012:10c-dsam,carlson2015:qmc-nuclear} (crosses)
  [off-scale in~(d) and~(f)], and the rotational $E2$ ratio (asterisk) are
  shown.  }
\label{fig:be2-norm-rp-scan-10be}
\end{figure*}

\subsection{Overview of results}
\label{sec:results-trans-survey:overview}

We turn now to transitions in nuclides where the ground state angular momentum
does not support a quadrupole moment [Fig.~\ref{fig:nuclear-chart-be2} (solid
  circles)].  We consider transitions from the first excited state to the ground
state~--- or, more generally, from the first excited state for which an $E2$
transition is allowed, by angular momentum and parity selection rules, to the
ground state: namely, $2^+\rightarrow0^+$ transitions in the mirror nuclides
$\isotope[10]{Be}$ and $\isotope[10]{C}$
(Sec.~\ref{sec:results-trans-survey:a10}) and in $\isotope[12]{C}$
(Sec.~\ref{sec:results-trans-survey:a12}), the $5/2^+\rightarrow1/2^+$
transition in $\isotope[11]{Be}$ (Sec.~\ref{sec:results-trans-survey:a11}), and
$3/2^-\rightarrow1/2^-$ transitions in the mirror nuclides $\isotope[13]{C}$ and
$\isotope[13]{N}$ (Sec.~\ref{sec:results-trans-survey:a13}).  The $E2$
strengths~\cite{npa1991:013-015,pritychenko2016:e2-systematics,dalessio2020:12c-escatt-be2}
are known except in $\isotope[11]{Be}$ and $\isotope[13]{N}$, while the
radii~\cite{angeli2013:charge-radii} are known except in $\isotope[10]{C}$ and
$\isotope[13]{N}$ (Fig.~\ref{fig:nuclear-chart-be2}).  For $\isotope[10]{Be}$,
where the strength of the transition from the second excited $2^+$ state to the
ground state is also experimentally known~\cite{mccutchan2009:10be-dsam-gfmc},
we consider this transition as well (Sec.~\ref{sec:results-trans-survey:a10}),
and, for $\isotope[12]{C}$, where the quadrupole moment of the excited $2^+$
state is experimentally
known~\cite{vermeer1983:12c-coulex,saizlomas2023:12c-coulex}, we consider this
quadrupole moment in relation to the ground state radius
(Sec.~\ref{sec:results-trans-survey:a12}).  We also make note of the calculated
dimensionless ratio $B(E2)/(eQ)^2$, involving the excited state quadrupole
moment, for these various transitions, as a diagnostic of axially symmetric
rotational structure and thus of the relevance of $B(E2)/(e^2r_p^4)$ to the
deformation, to which we return in Sec.~\ref{sec:results-trans-deformation}.

For a concise overview of the \textit{ab initio} results for $B(E2)/(e^2r_p^4)$,
in Fig.~\ref{fig:be2-norm-rp-ratio-teardrop-forbidden-q}, we restrict our
attention to a single, fixed value for the oscillator parameter $\hw$, namely,
$\hw=20\,\MeV$ (as in Figs.~\ref{part1:fig:q-norm-rp-ratio-teardrop-mirror}
and~\ref{part1:fig:q-norm-rp-ratio-teardrop-neqz} of Part~I), and examine the
$\Nmax$ dependence of the calculated results.  The results of three-point exponential
extrapolation (small circles) are again indicated in the case of the Daejeon16
interaction.  The experimental results are shown where available (horizontal
lines, with error bands), and GFMC AV18+IL7
predictions~\cite{pastore2013:qmc-em-alt9} (crosses) are shown for the $A=10$
nuclides~\cite{mccutchan2012:10c-dsam,carlson2015:qmc-nuclear}.
Comprehensive plots of the
calculated (and extrapolated) observables and ratios, as functions of both
$\Nmax$ and $\hw$, for both proton and neutron observables, are provided in the
Supplemental Material~\cite{supplemental-material}, along with numerical
tabulations of the calculated observables.

\subsection{$\isotope[10]{Be}$ \& $\isotope[10]{C}$}
\label{sec:results-trans-survey:a10}

Comparing the $A=10$ mirror nuclei, we may note that the convergence of
$B(E2)/(e^2r_p^4)$ is considerably slower for the proton-rich $\isotope[10]{C}$
[Fig.~\ref{fig:be2-norm-rp-ratio-teardrop-forbidden-q}(b)] than for the
neutron-rich $\isotope[10]{Be}$
[Fig.~\ref{fig:be2-norm-rp-ratio-teardrop-forbidden-q}(a)].  Relative
differences are about twice as large between the values for successive $\Nmax$
at the highest $\Nmax$ shown, but with similar difference ratios
($\eta\approx0.8$) for both nuclei.

In $\isotope[10]{Be}$, the first $2^+$ state lies $\approx3.5\,\MeV$ below the
neutron breakup threshold, and many $\MeV$ more below the proton threshold. In
the mirror nuclide $\isotope[10]{C}$, the first $2^+$ lies a mere
$\approx0.5\,\MeV$ below the proton breakup threshold~\cite{npa2004:008-010}.
One might therefore be tempted to attribute the difference in convergence
behavior just noted to threshold effects in $\isotope[10]{C}$.  However, the
calculations for the ratio $B(E2)/(e^2r_p^4)$ of proton obervables in
$\isotope[10]{C}$ and the corresponding ratio of neutron observables in $\isotope[10]{Be}$ are
mirror symmetric in their convergence behavior, to within minor quantitative
differences, as are the calculations for the proton ratio in $\isotope[10]{Be}$
and the corresponding neutron ratio in $\isotope[10]{C}$ (see Supplemental
Material~\cite{supplemental-material} for detailed convergence plots of these
ratios).  Thus, the
difference in convergence between the proton ratio in $\isotope[10]{Be}$
[Fig.~\ref{fig:be2-norm-rp-ratio-teardrop-forbidden-q}(a)] and the proton ratio
in $\isotope[10]{C}$ [Fig.~\ref{fig:be2-norm-rp-ratio-teardrop-forbidden-q}(b)]
largely reflects that between the (proton) $B(E2)/(e^2r_p^4)$ ratio in
$\isotope[10]{Be}$ and the corresponding neutron ratio in $\isotope[10]{Be}$
itself, which must be attributable to differences in the proton and neutron
structures within this nucleus, rather than differences in binding between the
mirror nuclei.  (Mirror symmetry is
considered again below in Sec.~\ref{sec:results-trans-deformation}.) 

The presumed $\alpha+\alpha+2n$ cluster molecular structure of the
$\isotope[10]{Be}$ ground state (\textit{e.g.},
Refs.~\cite{freer2006:10be-resonant-molecule,bohlen2007:10be-pickup,suzuki2013:6he-alpha-10be-cluster,kanadaenyo1999:10be-amd,li2023:p-palpha-molecular}),
together with the corresponding $\alpha+\alpha+2p$ structure of the
$\isotope[10]{C}$ mirror nucleus, provides an ostensible explanation for the
difference in convergence rates.  The $E2$ operator is only sensitive to the
protons.  In this picture, the protons in $\isotope[10]{Be}$ are all localized
within the $\alpha$ clusters, while, in $\isotope[10]{C}$, two of the protons
are in more diffuse molecular orbitals, which are more difficult to describe in
an oscillator basis.

In $\isotope[10]{Be}$, both the first and second $2^+$ states are
bound~\cite{npa2004:008-010}.  If one were to assume a simple, axially symmetric
rotational picture, the $2^+_1$ state (at $3.4\,\MeV$) would be a member of the
ground-state $K=0$ rotational band, while the $2^+_2$ state (at $6.0\,\MeV$)
would be understood to be the band head of a low-lying $K=2$ side band [see
  Fig.~1(a) of Ref.~\cite{caprio2019:bebands-sdanca19}].  However, the existence
of a low-lying $K=2$ band is suggestive of triaxial
structure~\cite{kanadaenyo1999:10be-amd,bohlen2007:10be-pickup,caprio2019:bebands-sdanca19,caprio2022:10be-shape-sdanca21},
which implies $K$ mixing~\cite{davydov1958:arm-intro}.  The $2^+_1\rightarrow
0^+_1$ strength is known from Doppler-shift lifetime
measurements~\cite{warburton1966:a10-a11-a12-dsam,fisher1968:a10-dsam,mccutchan2009:10be-dsam-gfmc},
which give
$B(E2;2^+_1\rightarrow0^+_1)=9.34({+46\atop-36})\,e^2\fm^4$~\cite{pritychenko2016:e2-systematics}.
For the $2^+_2\rightarrow 0^+_1$ transition, the $2^+_2$ lifetime is newly
measured by McCutchan \textit{et al.}~\cite{mccutchan2009:10be-dsam-gfmc}, since
the most recent evaluation~\cite{npa2004:008-010}, giving
$B(E2;2^+_2\rightarrow0^+_1)=0.11(2)\,e^2\fm^4$.  We consider convergence of
both the $2^+_1\rightarrow 0^+_1$ and $2^+_2\rightarrow 0^+_1$ transitions, and
the relevant dimensionless ratios, in calculations with the Daejeon16
interaction, in Fig.~\ref{fig:be2-norm-rp-scan-10be}.\footnote{Results for the
$2^+_1$ excitation energy and $2^+_1\rightarrow0^+_1$ $E2$ strength, with the
Daejeon16 interaction, were previously shown in Fig.~2 of
Ref.~\cite{caprio2019:bebands-sdanca19}.}

From GFMC calculations~\cite{mccutchan2009:10be-dsam-gfmc}, it has been observed
that these two $2^+$ states are subject to mixing, and that their energy
separation, and thus the actual degree of mixing, is highly sensitive to the
details of the internucleon interaction.  The calculated $E2$ strengths may then
be expected to be sensitive to such mixing.  In NCCI calculations, even for a
given interaction, the mixing is furthermore sensitive to the convergence of the
energy separation of the calculated levels with respect to the basis truncation.
Calculations with the Daejeon16 interaction give rapidly converging excitation
energies for both these $2^+$ states, as shown in
Fig.~\ref{fig:be2-norm-rp-scan-10be}(a), especially for the $2^+_1$ state,
mitigating such concerns.  These predicted excitation energies also agree with
experiment (solid squares) to well within half an $\MeV$.  However, in
calculations with the harder JISP16 and LENPIC interactions, the $2^+_2$
excitation energy is still poorly converged, and therefore so is the separation
in the $2^+$ energies.

The NCCI calculated $E2$ strength for the $2^+_1\rightarrow 0^+_1$ transition
[Fig.~\ref{fig:be2-norm-rp-scan-10be}(c)] shows some hints of shouldering, as
well as a crossing point for curves of successive $\Nmax$ at
$\hw\approx11\,\MeV\text{--}12\,\MeV$, in the general vicinity of the
experimental value.  Taking the dimensionless ratio $B(E2)/(e^2r_p^4)$
[Fig.~\ref{fig:be2-norm-rp-scan-10be}(e)] eliminates much of the $\hw$
dependence, and somewhat tames the $\Nmax$ dependence.  The ratio converges
steadily from below, much as seen above for $\isotope[9]{Be}$
[Fig.~\ref{fig:be2-norm-comparison-scan-9be}(j)], approaching both the
experimental ratio (solid square) and the GFMC AV18+IL7
prediction~\cite{mccutchan2012:10c-dsam,carlson2015:qmc-nuclear}~(cross).  An
exponential extrapolation for the ratio involving $B(E2;2^+_1\rightarrow0^+_1)$
[Fig.~\ref{fig:be2-norm-rp-scan-10be}(e) (dotted curves)] would place the
Daejeon16 prediction above the last calculated result ($\approx 0.35$) by
somewhere between once and twice again the size of last ``step'', and thus
generally within the experimental uncertainties.  (Calibrating to the
experimental ground state point-proton radius gives the scale shown at right.)

The dimensionless ratios, and thus estimated $B(E2)$, obtained with the JISP16
and LENPIC interactions [returning to
  Fig.~\ref{fig:be2-norm-rp-ratio-teardrop-forbidden-q}(a)] might superficially
appear to behave similarly to those for Daejeon16, but on closer examination the
LENPIC results do not clearly show a geometric progression towards convergence.
Although the calculated values at $\Nmax=12$ differ by $\lesssim5\%$, at this
scale, the differences in calculated values across these different interactions
cannot be distingished from the effects of incomplete convergence.

For the $2^+_2\rightarrow 0^+_1$ transition, the NCCI calculated $B(E2)$
[Fig.~\ref{fig:be2-norm-rp-scan-10be}(c)] has a more dramatic $\hw$-dependence
at low $\Nmax$, but the curves rapidly flatten for higher $\hw$.  In a
rotational picture, this transition is an interband transition, and its strength
might be expected to be strongly influenced by any small admixed contribution
from the ground state band's in-band $2\rightarrow0$ strength entering into the
interband strength, due to mixing of the two $2^+$ states.

Such dependence on mixing effects is not the type of systematic convergence behavior
which we might hope to offset by normalization to the radius.  Indeed, taking
the dimensionless ratio [Fig.~\ref{fig:be2-norm-rp-scan-10be}(f)] does not yield
any marked qualitative difference in the convergence pattern.  Nevertheless,
taking the ratio would appear (perhaps fortuitously) to provide some improvement
in the convergence behavior.  While increasing the spread of the curves for low
$\Nmax$ ($\Nmax\lesssim 6$), it flattens and compresses the curves obtained at
higher $\Nmax$.  We may thus read off an estimated ratio
$B(E2)/(e^2r_p^4)\approx 0.006$, or
$B(E2;2^+_2\rightarrow0^+_1)\approx0.14\,e^2\fm^4$ (scale at right).  The
predicted interband transition strength is roughly consistent with
experiment~(open square), although just outside the experimental uncertainties.
This agreement in scale is notable given that the $2^+_2\rightarrow 0^+_1$
interband transition is so comparatively weak: nearly two orders of magnitude
weaker than the collective in-band $2^+_1\rightarrow 0^+_1$ strength, and an
order of magnitude weaker than the Weisskopf estimate for a typical
single-particle strength in these nuclei ($\approx1.3\,e^2\fm^4$).

For the strong, in-band $2^+_1\rightarrow 0^+_1$ transition
[Fig.~\ref{fig:be2-norm-rp-scan-10be}(c)] the GFMC AV18+IL7
prediction~\cite{mccutchan2009:10be-dsam-gfmc,mccutchan2012:10c-dsam,carlson2015:qmc-nuclear}
(cross) is consistent with experiment, and in close agreement with the present
prediction for the dimensionless ratio
[Fig.~\ref{fig:be2-norm-rp-scan-10be}(e)].  However, the GFMC AV18+IL7
prediction for the weaker interband transition, at $1.7(1)\,e^2\fm^4$, lies off
scale in Fig.~\ref{fig:be2-norm-rp-scan-10be}(d), as does the corresponding
ratio in Fig.~\ref{fig:be2-norm-rp-scan-10be}(f).  Such a large difference in the calculated
strength of the weaker transition is perhaps not suprising, given the large
variations in mixing in the GFMC calculations~\cite{mccutchan2009:10be-dsam-gfmc}.

If we consider the quadrupole
moment of the $2^+_1$ state in relation to the $2^+_1\rightarrow0^+_1$
transition strength, the rotational model (with $K=0$) gives
$B(E2)/(eQ)^2=49/(64\pi)\approx0.2437$.  In the calculations with with Daejeon16
interaction, shown in
Fig.~\ref{fig:be2-norm-rp-scan-10be}(b),\footnote{Calculations of $B(E2)/(eQ)^2$
for $\isotope[10]{Be}$ and $\isotope[10]{C}$ with the Daejeon16 interaction, for
$\Nmax\leq10$, have previously been reported by Li \textit{et
  al.}~\cite{li2024:quadrupole-10be-c}.}  this ratio appears to be well
converged, but to a value $\approx20\%$ higher than the rotational
value.\footnote{Deviations from axially symmetric rotational relations within
the nominal $K=0$ ground state band, for both proton and neutron $E2$ matrix
elements, may already be seen in Fig.~13(b) of Ref.~\cite{maris2015:berotor2},
for calculations with the JISP16 interaction.}
Given the above discussion of mixing and triaxiality, we should perhaps not be suprised if the
calculated $2^+$ states are mixed relative to the pure rotational $K=0$ and $2$
states, leading to a breakdown of the axially symmetric rotational
picture.

\begin{figure*}
\centering
\includegraphics[width=\ifproofpre{0.8}{0.90}\hsize]{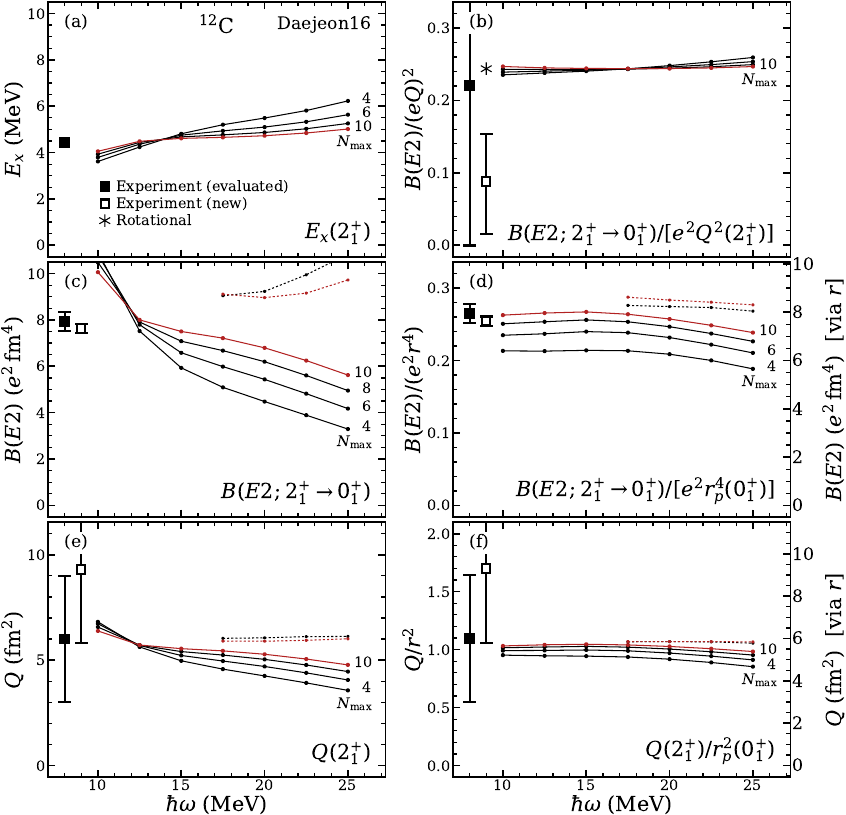}
\caption{Calculated observables for $\isotope[12]{C}$: (left) (a)~the $2^+$
  excitation energy, (c)~$B(E2;2^+_1\rightarrow 0^+_1)$, and (e)~the
  excited-state quadrupole moment $Q(2^+_1)$, and (right) dimensionless ratios
  constructed from these observables, including (b)~$B(E2)/(eQ)^2$, (d)~$B(E2)/(e^2r_p^4)$, taken
  relative to the ground state proton radius, and
  (f)~$Q/r_p^2$, again taken relative to the ground state radius.  Calculated
  values, for the Daejeon16 interaction, are shown as functions of the basis
  parameter $\hw$, for successive even values of $\Nmax$, from $\Nmax=4$ to $10$
  (as labeled).  When calibrated to the experimentally deduced value for $r_p$,
  the ratios provides predictions for the absolute $B(E2)$ and $Q$ (scale at
  right).  Exponential extrapolations (small circles, dotted lines) are
  provided, for selected observables.  For comparison, the experimental
  values~\cite{angeli2013:charge-radii,npa2017:012} are shown, taking either the
  evaluated $B(E2)$~\cite{pritychenko2016:e2-systematics} and
  $Q$~\cite{stone2016:e2-moments} (solid squares) or the more recent
  measurements of D'Alessio \textit{et al.}~\cite{dalessio2020:12c-escatt-be2}
  and Saiz-Lomas \textit{et al.}~\cite{saizlomas2023:12c-coulex}, respectively,
  for these observables (open squares), and the rotational $E2$ ratio (asterisk)
  is also shown.
}
\label{fig:be2-norm-rp-scan-12c}
\end{figure*}

\subsection{$\isotope[12]{C}$}
\label{sec:results-trans-survey:a12}

Moving on to $\isotope[12]{C}$, the $2^+_1$ state (at $4.4\,\MeV$) lies below
the $\alpha$ breakup threshold~\cite{npa2017:012} and is interpreted as a member
of the $K=0$ ground-state band.  We consider the $2^+_1\rightarrow 0^+_1$
transition [Fig.~\ref{fig:be2-norm-rp-ratio-teardrop-forbidden-q}(d)].  The
ratios $B(E2)/(e^2r_p^4)$ calculated with the JISP16 and LENPIC interactions, at
$\Nmax=10$, are $\approx10\text{--}20\%$ lower than obtained with the Daejeon16
interaction.  Although the convergence rates appear to be roughly similar for
these calculations, it is unclear how much of the difference in calculated
values might still be due to differences in convergence, rather than differences
in the true predictions with these interactions.

Let us examine the results of the calculations with the Daejeon16 interaction,
in Fig.~\ref{fig:be2-norm-rp-scan-12c}.  The excitation energy
[Fig.~\ref{fig:be2-norm-rp-scan-12c}(a)] is seen to be well converged and
consistent with experiment (solid square).

The NCCI calculated $E2$ strength [Fig.~\ref{fig:be2-norm-rp-scan-12c}(c)], again, shows some hints of shouldering and a crossing point (at
$\hw\approx11\,\MeV\text{--}12\,\MeV$), and these features are in the general
vicinity of the experimental strength.  In particular, the evaluated $E2$
strength is
$B(E2;2^+_1\rightarrow0^+_1)=7.9(4)\,e^2\fm^4$~\cite{pritychenko2016:e2-systematics}
(solid square), while the more recent electron scattering experiment of
D'Alessio \textit{et al.}~\cite{dalessio2020:12c-escatt-be2} yields a consistent
result, but with smaller uncertainties, of
$B(E2;2^+_1\rightarrow0^+_1)=7.63(19)\,e^2\fm^4$ (open square).  The results of
an exponential extrapolation (dotted curves) seem to be tending higher than
experiment, at $\approx9\,e^2\fm^4$, but not robustly.

Taking the dimensionless ratio $B(E2)/(e^2r_p^4)$
[Fig.~\ref{fig:be2-norm-rp-scan-12c}(d)] once again eliminates much of the $\hw$
dependence, while also modestly improving the $\Nmax$ dependence, thereby
permitting a more concrete comparison to be made.  The calculated ratios are
consistent in scale with the experimental results.  However, if we take the
evaluated $B(E2)$ (solid square), they are on track to exceed the upper end of
the experimental uncertainties, and, if we take D'Alessio \textit{et
  al.}\ result (open square), they already lie above the experimental
uncertainties, by $\Nmax=10$, in the vicinity of $\hw=15\,\MeV$.  The
exponential extrapolations (dotted curves) remove most of the remaining $\hw$
dependence but are still creeping upwards with increasing $\Nmax$.  Estimating
$B(E2)/(e^2r_p^4)\approx0.28$--$0.30$, gives an estimated $B(E2)$ similarly
tending above experiment, namely, $\approx8$--$9\,e^2\fm^4$ (scale at right).

Both in the context of rotational structure and for comparison with experiment,
it is of interest here to mention the quadrupole moment of the $2^+$ state
[Fig.~\ref{fig:be2-norm-rp-scan-12c}(e)].
The quadrupole moment is experimentally known, but with large uncertainties, from the
reorientation effect in Coulomb excitation~\cite{vermeer1983:12c-coulex}.  The
evaluated experimental quadrupole moment is
$Q(2^+_1)=+6(3)\,\fm^2$~\cite{stone2016:e2-moments} (solid square).  
However, the more recent experiment of Saiz-Lomas \textit{et
  al.}~\cite{saizlomas2023:12c-coulex} gives
$Q(2^+_1)=+9.3({+35\atop-38})\,\fm^2$ (open square).  While the uncertainties
overlap with the prior evaluated moment, the central value from the new
measurement is much larger.

The calculated ratio $B(E2;2^+_1\rightarrow0^+_1)/[eQ(2^+)]^2$
[Fig.~\ref{fig:be2-norm-rp-scan-12c}(b)]\footnote{Calculations of $B(E2)/(eQ)^2$
for $\isotope[12]{C}$ with the Daejeon16 interaction have also previously been
reported by Li \textit{et al.}~\cite{li2024:quadrupole-10be-c}.} closely
matches the rotational prediction
$B(E2;2\rightarrow0)/[eQ(2)]^2=49/(64\pi)\approx0.24$ (asterisk).  Calci and
Roth also verified this correlation, for a variety of NCCI calculations, in
Fig.~4 of Ref.~\cite{calci2016:observable-correlations-chiral} (see also Fig.~6
of Ref.~\cite{dalessio2020:12c-escatt-be2}).  The experimental uncertainties on
the evaluated quadrupole moment (solid square) leave the ratio $B(E2)/(eQ)^2$
poorly constrained (although the central value, in fact, happens to be closely
consistent with the rotational prediction).  However, the more recent
measurement (open square) is in strong tension with the rotational prediction,
and thus with the present NCCI predictions as well.  (See also Fig.~1 of
Ref.~\cite{saizlomas2023:12c-coulex}, which highlights the conflict between the
experimental result of that paper and various \textit{ab initio} and model
predictions.)

The calculated dimensionless ratio $Q/r_p^2$
[Fig.~\ref{fig:be2-norm-rp-scan-12c}(f)], involving the excited state quadrupole
moment but ground state radius, which may be used to estimate the quadrupole
moment by calibration to the ground state radius, has little freedom to surprise
us in its convergence behavior, given that we have already examined
$B(E2)/(e^2r_p^4)$ [Fig.~\ref{fig:be2-norm-rp-scan-12c}(d)]. That is to say,
since we have found the calculated $B(E2)/(eQ)^2$
[Fig.~\ref{fig:be2-norm-rp-scan-12c}(b)] to be essentially constant, the
calculated $Q/r_p^2$ must be simply proportional to the square root of the
calculated $B(E2)/(e^2r_p^4)$ [Fig.~\ref{fig:be2-norm-rp-scan-12c}(d)], so that
the overall behavior is the same, but with all relative differences
approximately halved.  Normalizing to the experimentally deduced
$r_p=2.340(3)\,\fm$~\cite{angeli2013:charge-radii} (scale at right) allows us to
estimate $Q(2^+_1)\approx6.0\,\fm^2$.  This finding is robust across choice of
interaction (see Fig.~3 of Ref.~\cite{caprio2022:emnorm}).  Here, again, if we
take the evaluated experimental quadrupole moment (solid square), the
predictions are near the central value, but, if we take the new
measurement~\cite{saizlomas2023:12c-coulex} (open square), the predictions are
at the lower edge of the uncertainties.

\begin{figure}
\centering
\includegraphics[width=\ifproofpre{0.9}{0.5}\hsize]{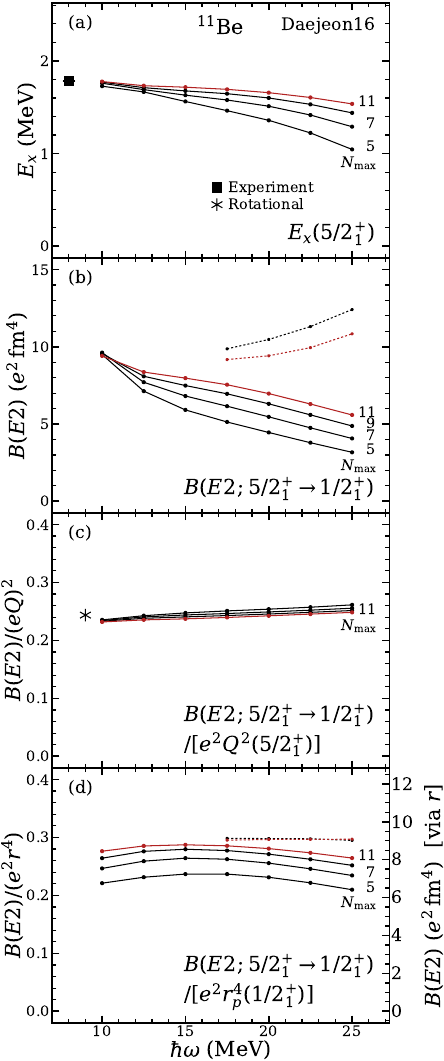}
\caption{Calculated observables for $\isotope[11]{Be}$: (a)~the $5/2^+_1$
  excitation energy, (b)~$B(E2; 5/2^+_1\rightarrow 1/2^+_1)$, (c)~the
  dimensionless ratio $B(E2)/(eQ)^2$, taken relative to the excited $5/2^+$
  state quadrupole moment, (d)~the dimensionless ratio $B(E2)/(e^2r_p^4)$, taken
  relative to the ground state radius. Calculated values, for the Daejeon16
  interaction, are shown as functions of the basis parameter $\hw$, for
  successive odd values of $\Nmax$, from $\Nmax=5$ to $11$ (as labeled).  When
  calibrated to the experimentally deduced value for $r_p$, the ratio provides a
  prediction for the absolute $B(E2)$ (scale at right).  Exponential
  extrapolations (small circles, dotted lines) are provided, for selected
  observables.  For comparison, the experimental excitation
  energy~\cite{npa2012:011} (square) is also shown.
}
\label{fig:be2-norm-rp-scan-11be}
\end{figure}

\subsection{$\isotope[11]{Be}$}
\label{sec:results-trans-survey:a11}

For $\isotope[11]{Be}$, we consider the $5/2^+\rightarrow1/2^+$ transition
[Fig.~\ref{fig:be2-norm-rp-ratio-teardrop-forbidden-q}(c)].  The $1/2^+$ ground
state~\cite{npa2012:011} is of non-normal parity~\cite{lane1960:reduced-widths},
\textit{i.e.}, opposite to that expected from simple sequential filling of
harmonic oscillator shells.  The $5/2^+$ state (at $1.8\,\MeV$), which lies above
the neutron breakup threshold and has a width of $\approx100\,\keV$, is the
first excited state of the same parity~\cite{npa2012:011}.  The convergence
patterns, for all three interactions, are qualitatively similar to those seen
above for neighboring $\isotope[10]{Be}$
[Fig.~\ref{fig:be2-norm-rp-ratio-teardrop-forbidden-q}(a)].  Any apparent
differences in calculated values across the three interactions, at finite
$\Nmax$, could well simply reflect the clearly incomplete convergence.

A closer examination of the convergence obtained with the Daejeon16 interaction,
in Fig.~\ref{fig:be2-norm-rp-scan-11be}, shows that, while three-point
exponential extrapolation is of little use for the $B(E2)$ itself
[Fig.~\ref{fig:be2-norm-rp-scan-11be}(b)], it is particularly robust, that is,
independent of $\Nmax$ and $\hw$, for the dimensionless ratio
[Fig.~\ref{fig:be2-norm-rp-scan-11be}(d)].  This is true at least for the
limited range of calculations available (from calculations with $\Nmax\geq5$, a
three-point extrapolation only becomes possible for
$\Nmax\geq9$).\footnote{States of non-normal parity are obtained from oscillator
configurations with an odd number of oscillator excitations relative to the lowest
Pauli-allowed filling of oscillator shells, and thus in odd-$\Nmax$ NCCI spaces.}  Calibration to the experimentally
deduced value of $r_p$ gives the scale at right in
Fig.~\ref{fig:be2-norm-rp-scan-11be}(d). Combining an extrapolated
$B(E2)/(e^2r_p^4)\approx0.30$, for the Daejeon16 interaction, with the
experimentally deduced $r_p=2.351(16)\,\fm$~\cite{angeli2013:charge-radii},
gives an estimated $B(E2;5/2^+\rightarrow1/2^+)\approx9.2\,e^2\fm^4$.

Note that the $5/2^+\rightarrow1/2^+$ transition, in a rotational picture, is
interpreted as a $K=1/2$ in-band transition [see Fig.~2(b) of
  Ref.~\cite{caprio2020:bebands}].  (Coriolis energy staggering raises the
$3/2+$ band member above the $5/2^+$ band member.) The calculated $B(E2)/(eQ)^2$
[Fig.~\ref{fig:be2-norm-rp-scan-11be}(c)], involving the excited $5/2^+$ state
quadrupole moment, is relatively well converged and consistent with the
rotational value $B(E2)/(eQ)^2=49/(64\pi)\approx0.2437$ for such a
transition.\footnote{Rough consistency with the axially symmetric rotational
relations within a $K=1/2$ band may already be seen for this transition in
Fig.~11(b) of Ref.~\cite{maris2015:berotor2}, for calculations with the JISP16
interaction.}

\begin{figure}
\centering
\includegraphics[width=\ifproofpre{0.9}{0.5}\hsize]{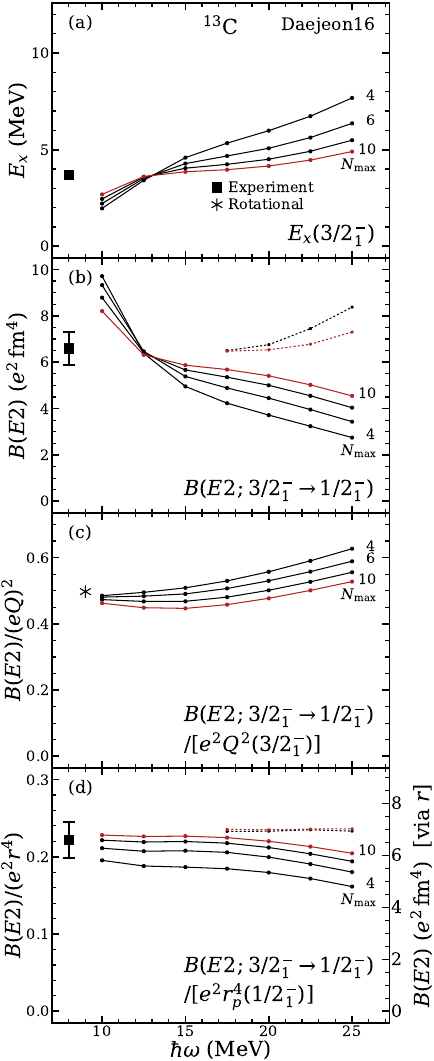}
\caption{Calculated observables for $\isotope[13]{C}$: (a)~the $3/2^-_1$
  excitation energy, (b)~$B(E2; 3/2^-_1\rightarrow 1/2^-_1)$,
  (c)~the
  dimensionless ratio $B(E2)/(eQ)^2$, taken relative to the excited $3/2^-$ state quadrupole moment,
  and (d)~the
  dimensionless ratio $B(E2)/(e^2r_p^4)$, taken relative to the ground state
  radius. Calculated values, for the Daejeon16 interaction, are shown as
  functions of the basis parameter $\hw$, for successive even values of $\Nmax$,
  from $\Nmax=4$ to $10$ (as labeled).  When calibrated to the experimentally
  deduced value for $r_p$, the ratio provides a prediction for the absolute
  $B(E2)$ (scale at right).  Exponential extrapolations (small circles, dotted
  lines) are provided, for selected observables.  For comparison, experimental
  values~\cite{npa1991:013-015,angeli2013:charge-radii} (squares) are also
  shown.  }
\label{fig:be2-norm-rp-scan-13c}
\end{figure}

\subsection{$\isotope[13]{C}$ \& $\isotope[13]{N}$}
\label{sec:results-trans-survey:a13}

Finally, let us consider the $3/2^-\rightarrow1/2^-$ transitions in the $A=13$
mirror nuclei, $\isotope[13]{C}$
[Fig.~\ref{fig:be2-norm-rp-ratio-teardrop-forbidden-q}(e)] and $\isotope[13]{N}$
[Fig.~\ref{fig:be2-norm-rp-ratio-teardrop-forbidden-q}(f)].
In $\isotope[13]{C}$, the $3/2^-$ excited state is bound, while, in the
mirror nuclide $\isotope[13]{N}$, this excited state lies above the proton
breakup threshold, and has a width of $\approx64\,\keV$~\cite{npa1991:013-015}.

The convergence rates for $B(E2)/(e^2r_p^4)$ are generally similar in these two
cases.  However, the convergence is less smooth for $\isotope[13]{N}$, at least
for the Daejeon16 calculations, where the values at higher $\Nmax$ become
compressed (in fact, the value obtained for $\Nmax=10$ lies marginally
\textit{below} that for $\Nmax=8$).  The calculated ratios, at finite $\Nmax$,
are higher for $\isotope[13]{C}$ than for $\isotope[13]{N}$, but, again,
different convergence rates can render such naive comparison misleading.

The convergence of the Daejeon16 results for $\isotope[13]{C}$ is shown in more
detail in Fig.~\ref{fig:be2-norm-rp-scan-13c}.  The excitation energy for the
$3/2^-$ state [Fig.~\ref{fig:be2-norm-rp-scan-13c}(a)] is consistent with
experiment.  The calculated $B(E2)$ values
[Fig.~\ref{fig:be2-norm-rp-scan-13c}(b)] are on the same scale as
experiment~\cite{npa1991:013-015}, and are plausibly consistent (\textit{e.g.},
from the crossing points of the curves or taking the exponential extrapolations
as a crude guide), but cannot be compared quantitatively.

Turning to the ratio $B(E2)/(e^2r_p^4)$
[Fig.~\ref{fig:be2-norm-rp-scan-13c}(d)], the calculated values are relatively
flat with respect to $\hw$, and the spacing is decreasing rapidly with $\Nmax$
($\eta\approx0.6$ at $\hw=20\,\MeV$).  The extrapolations obtained at $\Nmax=8$
and $\Nmax=10$ are nearly indistinguishable and independent of $\hw$, suggesting
a ratio of $B(E2)/(e^2r_p^4)\approx 0.23$--$0.24$.  This lies just within the
uncertainties of the experimental value
$B(E2)/(e^2r_p^4)=0.22(2)$~\cite{angeli2013:charge-radii,npa1991:013-015}.

Note that the $3/2^-\rightarrow1/2^-$ transition, in a rotational picture, is
most simply interpreted as a $K=1/2$ in-band transition.  The calculated
$B(E2)/(eQ)^2$ [Fig.~\ref{fig:be2-norm-rp-scan-13c}(c)], involving the excited
$3/2^-$ state quadrupole moment, is not as well converged as in some of the
previous examples, but is roughly consistent with the rotational value
$B(E2)/(eQ)^2=25/(16\pi)\approx0.4947$ for such a transition.

 \begin{figure*}
\centering
\includegraphics[width=\ifproofpre{0.70}{1.0}\hsize]{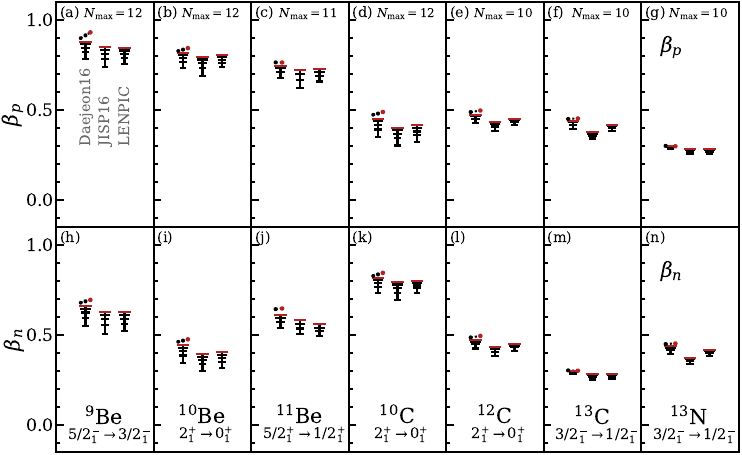}
\caption{Calculated proton~(top) and neutron~(bottom) deformations, as deduced
  from the ratios of the form $B(E2)/(e^2r^4)$, under the assumption of axially
  symmetric rotation, for the nuclides considered in this work.  Results are obtained with the Daejeon16,
  JISP16, and LENPIC interactions (from left to right, within each panel).
  Calculated values are shown at fixed $\hw=20\,\MeV$ and varying $\Nmax$
  (increasing tick size), from $\Nmax=4$ (or $5$) to the maximum value indicated
  (at top).  The initial and final states for the transition used are indicated
  (at bottom), with $K$ as noted in the text.}
\label{fig:beta-from-ratio-be2-r4-teardrop}
\end{figure*}

\section{Deformation}
\label{sec:results-trans-deformation}

Finally, let us explore what the ratios of the form $B(E2)/(e^2r^4)$ indicate for the
nuclear quadrupole deformation, via the rotational
relation
given in~(\ref{part1:eqn:be2-norm-rp-beta-proton}) of Part~I.  That is, for the proton observables,
\begin{equation*}
\frac{B(E2;J_i\rightarrow J_f)}{Z^2e^2r_p^4}=\Bigl(\frac{5}{4\pi}\Bigr)^2
  \tcg{J_i}{K}{2}{0}{J_f}{K}^2 \beta_p^2,
\end{equation*}
where $K$ is the angular momentum projection of the rotational intrinsic state
along the symmetry axis (see Sec.~\ref{part1:sec:background:deformation} of
Part~I).  The analogous ratio $B(E2_n)/(e^2r_n^4)$, \textit{i.e.}, calculated
using the neutron quadrupole operator and neutron radius, similarly yields the
neutron deformation $\beta_n$, with the replacement of $Z$ by $N$ in the above
expression.  We recall here that ratios of the form $B(E2)/(e^2r^4)$, like
$\beta$ itself, are unsigned, and do not distinguish between prolate and oblate
deformation.  Furthermore, while~(\ref{part1:eqn:be2-norm-rp-beta-proton}) of
Part~I is applicable either to prolate or oblate deformation, it does assume
axial symmetry.  The resulting deformations thus extracted from the transitions
considered in this work are shown in
Fig.~\ref{fig:beta-from-ratio-be2-r4-teardrop}, for both the protons~(top) and
neutrons~(bottom).

Here it is worth highlighting that the traditional expression
commonly used to extract the nuclear deformation from the $0^+\rightarrow 2^+$
$E2$ strength in empirical tabulations for even-even
nuclei~\cite{raman2001:systematics,pritychenko2016:e2-systematics},
reproduced
in~(\ref{part1:eqn:beta-be2-norm-r-proton-k0-traditional}) of Part~I, namely,
\begin{equation*}
\beta = \frac{4\pi}{3}\frac{B(E2;0\rightarrow2)^{1/2}}{Zer_0^2A^{2/3}},
\end{equation*}
is deduced from this same relation, but relies upon the textbook global
phenomenological estimate for the radius ($r\propto A^{1/3}$).  This leads to
notable discrepancies for the nuclei being considered here.

For $\isotope[10]{Be}$, the proton radius deduced from the measured charge
radius~\cite{angeli2013:charge-radii} (see Sec.~\ref{part1:sec:background:rc} of
Part~I) is $\approx11\%$ higher than the global phenomenological estimate, or,
for $\isotope[12]{C}$, $\approx10\%$ higher.  This radius then enters
quadratically into the determination of the $\beta$, doubling the effect (to
$\approx20\%$).  Thus, while the calculated $B(E2)/(e^2r_p^4)$ is in general
agreement with experiment for $\isotope[10]{Be}$
[Fig.~\ref{fig:be2-norm-rp-ratio-teardrop-forbidden-q}(a)] and $\isotope[12]{C}$
[Fig.~\ref{fig:be2-norm-rp-ratio-teardrop-forbidden-q}(d)], and thus the deduced
(proton) deformations in Fig.~\ref{fig:beta-from-ratio-be2-r4-teardrop}(b,e) are
in similar agreement with those deduced from the experimental ratio, the
tabulated empirical deformations obtained from the very same measured $E2$
strengths~\cite{pritychenko2016:e2-systematics}, but relying upon the global
empirical fit estimate for the radius, are higher: $\beta\approx1.07$ for
$\isotope[10]{Be}$ [Fig.~\ref{fig:beta-from-ratio-be2-r4-teardrop}(b)] and
$\beta\approx0.58$ for $\isotope[12]{C}$
[Fig.~\ref{fig:beta-from-ratio-be2-r4-teardrop}(e)].  The tabulated deformation
of $\beta\approx0.70$ for $\isotope[10]{C}$
[Fig.~\ref{fig:beta-from-ratio-be2-r4-teardrop}(d)] is likewise higher than
calculated, but here the experimental charge radius is not
known~\cite{angeli2013:charge-radii} to permit comparison with the experimental
ratio.

We may also note that mirror symmetry holds to a very good approximation for the
calculated ratios $B(E2)/(e^2r^4)$, and thus for the extracted deformations in
Fig.~\ref{fig:beta-from-ratio-be2-r4-teardrop}, even though the \textit{ab
  initio} calculations allow for isospin symmetry breaking (through the Coulomb
interaction between protons in the Daejeon16 and JISP16 calculations and through
additional isospin symmetry breaking contributions to the nuclear force in the
LENPIC calculations).  In particular, the extracted deformations for the mirror
nuclei $\isotope[10]{Be}$ [Fig.~\ref{fig:beta-from-ratio-be2-r4-teardrop}(b,i)]
and $\isotope[10]{C}$ [Fig.~\ref{fig:beta-from-ratio-be2-r4-teardrop}(d,k)] are
nearly indistinguishable, under interchange of the proton and neutron
deformations, and similarly for $\isotope[13]{C}$
[Fig.~\ref{fig:beta-from-ratio-be2-r4-teardrop}(f,m)] and $\isotope[13]{N}$
[Fig.~\ref{fig:beta-from-ratio-be2-r4-teardrop}(g,n)].  Likewise, the calculated
proton and neutron deformations for the $N=Z$ nuclide $\isotope[12]{C}$
[Fig.~\ref{fig:beta-from-ratio-be2-r4-teardrop}(e,l)] are nearly
indistinguishable.

\begin{figure*}
\centering
\includegraphics[width=\ifproofpre{0.70}{1.0}\hsize]{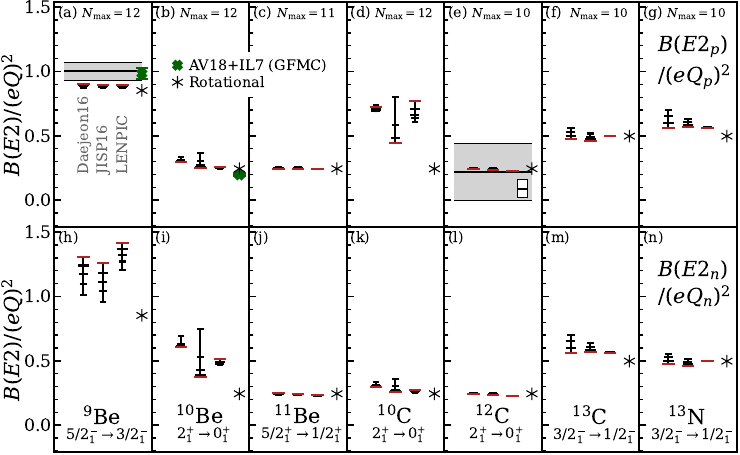}
\caption{Calculated ratios $B(E2)/(eQ)^2$ obtained for the proton~(top) and
  neutron~(bottom) quadrupole operators, for the nuclides considered in this
  work.  Results are obtained with the Daejeon16, JISP16, and LENPIC
  interactions (from left to right, within each panel).  Calculated values are
  shown at fixed $\hw=20\,\MeV$ and varying $\Nmax$ (increasing tick size), from
  $\Nmax=4$ (or $5$) to the maximum value indicated (at top).  The initial and
  final states for the transition used are indicated (at bottom), with $Q$ taken
  from the ground state for $\isotope[9]{Be}$ and excited state otherwise.  For
  comparison, the experimental
  ratios~\cite{angeli2013:charge-radii,pritychenko2016:e2-systematics,npa1991:013-015}
  are shown (horizontal line and error band), as are the GFMC AV18+IL7
  predictions~\cite{pastore2013:qmc-em-alt9,mccutchan2012:10c-dsam,carlson2015:qmc-nuclear}
  [off-scale for $\isotope[10]{C}$ in~(d)] (crosses), and rotational $E2$ ratios
  (asterisks).  For $\isotope[12]{C}$, the ratio deduced from the more recent
  measurements of D'Alessio \textit{et al.}~\cite{dalessio2020:12c-escatt-be2}
  and Saiz-Lomas \textit{et al.}~\cite{saizlomas2023:12c-coulex} is also
  included (horizontal line and error band, shown as narrow open box).  }
\label{fig:be2-norm-q-both-species-teardrop}
\end{figure*}

For the $\isotope{Be}$ isotopes, in the cluster molecular orbital
interpretation~\cite{hiura1963:alpha-model-9be,*hiura1964:alpha-model-9be-ERRATUM,okabe1979:9be-molecular-part-1,*okabe1979:9be-molecular-part-2,seya1981:molecular-orbital,vonoertzen1996:be-molecular,vonoertzen1997:be-alpha-rotational,freer2007:cluster-structures},
supported by microscopic antisymmetrized molecular dynamics
(AMD)~\cite{kanadaenyo1997:c-amd-pn-decoupling,kanadaenyo1999:10be-amd,suhara2010:amd-deformation}
and nuclear lattice effective field theory (NLEFT)
calculations~\cite{shen2025:be-nleft}, as well as NCCI
results~\cite{maris2012:mfdn-ccp11}, the nucleus consists of an underlying
$\alpha+\alpha$ dimer ($\isotope[8]{Be}$), with the additional ``valence''
neutrons occupying $\pi$ (equatorial) or $\sigma$ (polar) molecular orbitals.
We might naively expect the proton deformation to be unchanged by the addition
of spectator neutrons.  However, even within the cluster molecular orbital
description this cannot be taken as a given: the presence of the valence
neutrons can affect the inter-$\alpha$ spacing, or perturb the structure of the
$\alpha$ particles, or induce recoil effects.\footnote{The proton and quadrupole
and monopole operators entering into the definition of the deformation for this
distribution (Sec.~\ref{part1:sec:background:deformation} of Part~I), are
defined relative to the center of mass of the combined proton and neutron
system~\cite{caprio2020:intrinsic}.  The neutron coordinates enter implicitly
into these observables through their effect on the center of mass, and the
addition of neutrons can affect proton observables by perturbing the center of
mass, in a so-called recoil effect.  As a prominent example, the enhanced proton
radius in the neutron halo nucleus $\isotope[6]{He}$ is understood to arise from
such a recoil effect~\cite{lu2013:laser-neutron-rich}.}  In the cluster
molecular orbital picture, the valence neutron in the $\isotope[9]{Be}$ ground
state occupies a $\pi$ equatorial orbital, which would be expected to lead to a
more spherical overall distribution for the neutrons than for the protons.  For
the $\isotope[10]{Be}$ ground state, both neutrons occupy $\pi$ orbitals.  The
AMD results indicate their influence is to induce an oblate deformation for the
neutrons, which aligns orthogonally to the prolate proton deformation, yielding
an overall triaxial deformation.  The positive parity (parity-inverted) ground
state of $\isotope[11]{Be}$ then is obtained from a $\pi^2\sigma$ configuration
of neutrons.

If we take the extracted deformations for the $\isotope{Be}$ isotopes in
Fig.~\ref{fig:beta-from-ratio-be2-r4-teardrop} at face value, the proton
deformation $\beta_p$ [Fig.~\ref{fig:beta-from-ratio-be2-r4-teardrop}(a,b,c)] is
not strictly constant across the nuclides, but demonstrates a gentle downward
drift from $\isotope[9]{Be}$ to $\isotope[11]{Be}$, with the calculated values
declining from $\approx0.8$--$0.9$ to $\approx0.7$--$0.8$.  (The same caveats
regarding incomplete convergence as discussed for the underlying ratio in
Sec.~\ref{sec:results-trans-survey} apply here as well.)  The neutron
deformation $\beta_n$ [Fig.~\ref{fig:beta-from-ratio-be2-r4-teardrop}(h,i,j)] is
more markedly variable, first descending from $\approx0.6$--$0.7$ to
$\approx0.4$--$0.5$, then rising again to $\approx0.6$.

However, it is important to keep in mind
that~(\ref{part1:eqn:be2-norm-rp-beta-proton}) of Part~I is based on the
assumption both of a rigid intrinsic deformation and of axially symmetric
rotational relations for $E2$ matrix elements.  The deformation extracted from
an $E2$ transition [via the ratio of the form $B(E2)/(e^2r^4)$,
  by~(\ref{part1:eqn:be2-norm-rp-beta-proton}) of Part~I] or from a quadrupole
moment [via the ratio of the form $Q/r^2$,
  by~(\ref{part1:eqn:Q-norm-r-beta-proton}) of Part~I] will be consistent only
if $B(E2)/(eQ)^2$ takes on its rotational value.  We have already made
preliminary comparisons of $B(E2)/(eQ)^2$ to the rotational ratio for several of
the transitions discussed above (Secs.~\ref{sec:results-trans-9be}
and~\ref{sec:results-trans-survey}).  A more comprehensive overview of
$B(E2)/(eQ)^2$, for all three interactions considered, and for both proton~(top)
and neutron~(bottom) quadrupole observables, is provided in
Fig.~\ref{fig:be2-norm-q-both-species-teardrop}.

Even for $\isotope[9]{Be}$, where we saw in Sec.~\ref{sec:results-trans-9be}
that the proton $B(E2)/(eQ)^2$
[Fig.~\ref{fig:be2-norm-q-both-species-teardrop}(a)] is relatively
well-converged and lies not far above the rotational ratio, the analogous ratio
for the neutron observables
[Fig.~\ref{fig:be2-norm-q-both-species-teardrop}(h)], is egregiously
unconverged.  Furthermore, by $\Nmax=12$ (at $\hw=20\,\MeV$), it is a factor of
$\approx1.5$ above the rotational ratio, and tending higher.  Correspondingly,
the extracted neutron deformations, at $\beta_n\gtrapprox0.6$, are tending
higher than the value $\beta_n\approx0.5$ obtained from the ground state
quadrupole moment (Fig.~\ref{part1:fig:beta-from-ratio-q-rsqr-scan-9be} of
Part~I) and, given the breakdown of the rotational picture, should not be taken
to be meaningful.  One may speculate that this poor convergence of the neutron
ratio is due to the behavior of the last neutron, which is bound in the
$3/2^-$ ground state, but unbound in the $5/2^-$ initial state for the
transition~\cite{npa2004:008-010}.

Then, recall from Sec.~\ref{sec:results-trans-survey:a10} that
$\isotope[10]{Be}$ exhibits blatant deviations from the rotational value for
$B(E2)/(eQ)^2$ [Fig.~\ref{fig:be2-norm-q-both-species-teardrop}(b)], by
$\approx20\%$, for the Daejeon16 interaction, and is suspected of triaxial,
rather than axially symmetric, structure.  The results for $B(E2)/(eQ)^2$, for
the JISP16 and LENPIC interactions, actually match the rotational prediction
more closely.  The corresponding neutron ratios
[Fig.~\ref{fig:be2-norm-q-both-species-teardrop}(i)] are slow to converge and
highly dependent upon interaction, but they are generally well above with the
rotational ratio, indeed, in the Daejeon16 results, $\gtrsim2.5$ times the
rotational value.  Thus, for $\isotope[10]{Be}$, as for $\isotope[9]{Be}$ above,
the neutron deformation extracted by application of the rotational relations is
not meaningful, nor, given the overall deviations from axially symmetric
rotational structure, would we have confidence in the proton deformation thus
extracted.

For $\isotope[11]{Be}$, the $5/2^+$ initial state for the
$5/2^+\rightarrow1/2^+$ transition is likewise unbound against neutron
emission~\cite{npa2012:011}, as for $\isotope[9]{Be}$ above.  Nonetheless, in
this case the calculated $B(E2)/(eQ)^2$ is well-converged both for the protons
[Fig.~\ref{fig:be2-norm-q-both-species-teardrop}(c)] and for the neutrons
[Fig.~\ref{fig:be2-norm-q-both-species-teardrop}(j)], and closely agrees with
the rotational ratio, across all three interactions.  Thus, among the
$\isotope{Be}$ isotopes considered here, only for $\isotope[11]{Be}$ would it
seem reasonable to attach meaning to the deformations extracted in
Fig.~\ref{fig:beta-from-ratio-be2-r4-teardrop}.  Note the lower overall
deformation for the neutrons [Fig.~\ref{fig:beta-from-ratio-be2-r4-teardrop}(j)]
than for the protons [Fig.~\ref{fig:beta-from-ratio-be2-r4-teardrop}(c)],
ostensibly attributable to the net result of adding equatorial ($\pi$) and polar
($\sigma$) valence neutrons.

Turning to the $\isotope{C}$ isotopes, $\isotope[10]{C}$
[Fig.~\ref{fig:beta-from-ratio-be2-r4-teardrop}(d,k)] is the mirror nuclide to
$\isotope[10]{Be}$, so there are similarly deviations from axially symmetric
rotation,\footnote{Intriguingly, we may observe significant deviations from
mirror symmetry in the calculated
$B(E2)/(eQ)^2$ for $\isotope[10]{Be}$ and $\isotope[10]{C}$, under interchange
of proton and neutron observables, if we compare
Fig.~\ref{fig:be2-norm-q-both-species-teardrop}(d) with
Fig.~\ref{fig:be2-norm-q-both-species-teardrop}(i) (see also Supplemental
Material~\cite{supplemental-material} for detailed convergence plots).} and it is again not
meaningful to extract a deformation from the $B(E2)$.  However, for
$\isotope[12]{C}$, recall from Sec.~\ref{sec:results-trans-survey:a12} that
$B(E2)/(eQ)^2$ [Fig.~\ref{fig:be2-norm-q-both-species-teardrop}(e,l)] is
robustly rotational.  For $\isotope[13]{C}$, as noted in
Sec.~\ref{sec:results-trans-survey:a13}, this ratio is less well converged, but
generally consistent with the rotational value, which we find across
interactions for both the proton
[Fig.~\ref{fig:be2-norm-q-both-species-teardrop}(f)] and neutron
[Fig.~\ref{fig:be2-norm-q-both-species-teardrop}(m)] ratios.

Thus, we may attach some credence to the extracted deformations for
$\isotope[12]{C}$ [Fig.~\ref{fig:beta-from-ratio-be2-r4-teardrop}(e,l)] and
$\isotope[13]{C}$ [Fig.~\ref{fig:beta-from-ratio-be2-r4-teardrop}(f,m)].  The
extracted proton deformations ($\approx0.4$--$0.5$) are essentially
indistinguishable across these two nuclei, while, going from $\isotope[12]{C}$
to $\isotope[13]{C}$, the neutron deformation drops (from $\approx0.4$--$0.5$,
the same as for the protons in $\isotope[12]{C}$) to $\approx0.3$, consistently
across the different interactions considered.

\section{Conclusion}
\label{sec:concl}

Although $E2$ transition strengths are often poorly convergent in NCCI
calculations, it is already known that correlations with other calculated
observables~\cite{calci2016:observable-correlations-chiral,caprio2022:8li-trans}
can be exploited to extract meaningful predictions.  It is most natural and
obvious to make of use correlations among the $E2$ observables themselves,
\textit{e.g.}, predicting an $E2$ strength by calibration to a known ground
state quadrupole moment, where such an approach is possible.  However, we
demonstrate that correlations between the calculated $E2$ strengths and the
ground state radius can also be exploited.  In the previous article (Part~I), we
assessed correlations between the $E2$ moment and the radius, through the
dimensionless ratio $Q/r_p^2$.  In the present article (Part~II), we explored
correlations between low-lying $E2$ transition strengths and the radius, with
the goal of using the dimensionless ratio $B(E2)/(e^2r_p^4)$ to predict an $E2$
strength by calibration to a known ground state radius.

Correlations between $E2$ strengths and the radius are not found to be as robust
as those involving only $E2$ observables, as seen in direct comparisons of the
dimensionless ratios $B(E2)/(eQ)^2$ and $B(E2)/(e^2r_p^4)$ for the same
transitions (\textit{e.g.}, Fig.~\ref{fig:be2-norm-comparison-scan-9be}).
Nonetheless, calibration to the ground state radius in many cases provides at
least a rough prediction where none might otherwise be available, as we
demonstrate for several $p$-shell nuclei for which the ground state angular
momentum does not admit a quadrupole moment
(Fig.~\ref{fig:be2-norm-rp-ratio-teardrop-forbidden-q}).  Furthermore,
\textit{ab initio} calculations for $B(E2)/(e^2r_p^4)$ provide an indirect
measure of the predicted deformation for these nuclei, under the assumption of
axially symmetric rotation.

It is worth noting some \textit{a priori} limitations to an approach which
relies upon robust prediction of correlations between an $E2$ transition
strength and a ground state property, whether the quadrupole moment or the
radius.  Such approaches are geared towards eliminating smoothly varying
systematic truncation error, rather than sudden changes arising from level
crossings and the consequent two-state mixing~\cite{casten2000:ns}.  Thus, in particular, they are
not immediately applicable to transitions between normal and intruder states,
which arise primarily from such mixing (\textit{e.g.},
Ref.~\cite{mccoy2024:12be-shape}).  Moreover, the convergence properties of
normal and intruder states in NCCI calculations are notably different
(\textit{e.g.}, Ref.~\cite{caprio2020:bebands}), so predictions of $E2$
properties of the intruder states themselves, or transitions among them, are not
likely candidates for calibration to ground state properties.

Nonetheless, we demonstrate that normalizing an $E2$ observable to the
appropriate power of the radius, to yield a dimensionless ratio, tames the $\hw$
dependence of the results and, to a less dramatic extent, the $\Nmax$ dependence
of the results.  We have explored the properties of such dimensionless ratios
and their predictive value, for transitions between low-lying states and the
ground state, for nuclei across the $p$ shell.

\begin{acknowledgments}
  We thank James P.~Vary, Ik Jae Shin, and Youngman Kim for sharing illuminating
  results on ratios of observables, Augusto O.~Macchiavelli for valuable
  discussions, and Scott R.~Carmichael, Johann Isaak, and Shwetha L.~Vittal for
  comments on the manuscript.  This material is based upon work supported by the
  U.S.~Department of Energy, Office of Science, under Awards
  No.~DE-FG02-95ER40934, DE-AC02-06CH11357, and DE-SC0023495 (SciDAC5/NUCLEI).
  This research used resources of the National Energy Research Scientific
  Computing Center (NERSC), a DOE Office of Science User Facility supported by
  the Office of Science of the U.S.~Department of Energy under Contract
  No.~DE-AC02-05CH11231, using NERSC awards NP-ERCAP0020422, NP-ERCAP0023497,
  and NP-ERCAP0026449.
\end{acknowledgments}

\bibliographystyle{apsrev4-2}
\nocite{control:title-on}

\end{document}